\documentclass[apj]{emulateapj}
\slugcomment{{\sc Accepted to Icarus: 17 May 2013}} 

\usepackage{graphics}
\usepackage{natbib}
\usepackage{multirow}
\usepackage{rotating}
\usepackage{amsfonts}
\usepackage{amsmath}
\usepackage{enumerate}
\usepackage{tabularx}
\usepackage{natbib,amsmath,amssymb,epsfig,graphicx,longtable,multirow,float,url,footnote}

\newcommand{\CLEAN}{{\tt CLEAN}}

\usepackage{url}
\usepackage{epstopdf}

\begin{document}
\title{Spatially-Resolved Millimeter-Wavelength Maps of Neptune }

\shortauthors{S. H. Luszcz-Cook, I. de Pater, M. Wright}

\author{
  S. H. Luszcz-Cook\altaffilmark{1}\altaffilmark{2},
  I. de Pater\altaffilmark{1}\altaffilmark{3},
    M. Wright \altaffilmark{1}
}

\altaffiltext{1}{Astronomy Department, University of California, 
Berkeley, CA 94720, USA}
\altaffiltext{2}{Astrophysics Department, American Museum of Natural History, Central Park West at 79th Street, New York, NY 10024, USA; shcook@amnh.org}
\altaffiltext{3}{SRON Netherlands Institute for Space Research, 3584 CA Utrecht, and Faculty of Aerospace Engineering, Delft University of Technology, 2629 HS Delft, The Netherlands}

\begin{abstract}
We present maps of Neptune in and near the CO (2-1) rotation line at 230.538 GHz. These data, taken with the Combined Array for Research in Millimeter-wave Astronomy (CARMA) represent the first published spatially-resolved maps in the millimeter. At large ($\sim$ 5 GHz) offsets from the CO line center, the majority of the emission originates from depths of 1.1--4.7 bar. We observe a latitudinal gradient in the brightness temperature at these frequencies, increasing by 2--3 K from 40$^\circ$N to the south pole. This corresponds to a decrease in the gas opacity of about 30\% near the south pole at altitudes below 1 bar, or a decrease of order a factor of 50 in the gas opacity at pressures greater than 4 bar. We look at three potential causes of the observed gradient:  variations in the tropospheric methane abundance, variations in the H$_2$S abundance, and deviations from equilibrium in the {\it ortho/para} ratio of hydrogen. At smaller offsets (0--213 MHz) from the center of the CO line, lower atmospheric pressures are probed, with contributions from mbar levels down to several bars. We find evidence of latitudinal variations at the 2-3\% level in the CO line, which are consistent with the variations in zonal-mean temperature near the tropopause found by \cite{conrath98} and \cite{orton07}.

\vspace{\baselineskip}
\small{NOTICE: this is the authors' version of a work that was accepted for publication in Icarus. Changes resulting from the publishing process may not be reflected in this document.}

\end{abstract}


 
\section{Introduction}\label{sec:cointro}

Neptune's millimeter continuum originates in the troposphere, from pressures of 1--5 bar. While collision-induced absorption of H$_2$ with hydrogen, helium and methane dominates the opacity at these wavelengths, several trace species also contribute, particularly H$_2$S, PH$_3$ and NH$_3$. The abundances of these trace species have yet to be uniquely determined, though good fits to centimeter-wavelength disk-integrated spectra, which probe depths of several bars down to 10's of bars, are obtained using an H$_2$S abundance that represents a 30--50 times enrichment  above the protosolar S/H value \citep{depater91,deboer96} and a protosolar abundance or less of  nitrogen in NH$_3$ \citep{romani89,depater91}. 

Another trace atmospheric species, carbon monoxide (CO), produces strong rotational lines at (sub)millimeter wavelengths.  CO is present in Neptune's upper atmosphere at a level several orders of magnitude greater than expected under thermochemical equilibrium conditions  \citep{marten91}. Two major pathways have been identified for enriching Neptune's atmosphere in CO, and these pathways result in different vertical CO abundance profiles. The first of these is upward mixing of CO from warm, deep layers of the atmosphere where CO is thermochemically stable \citep{lodders94}; CO supplied in this fashion will be well-mixed throughout the upper atmosphere and therefore will exhibit a uniform vertical profile. Alternatively, CO may be produced in the stratosphere of the planet as a result of the infall of oxygen-bearing material; in this case, downward transport will act as a sink and the CO abundance will fall below observable levels in the troposphere, where the diffusion rate increases dramatically \citep{moses92}.  
\begin{figure*}[t]
\begin{center}$
\includegraphics[width=0.4\textwidth]{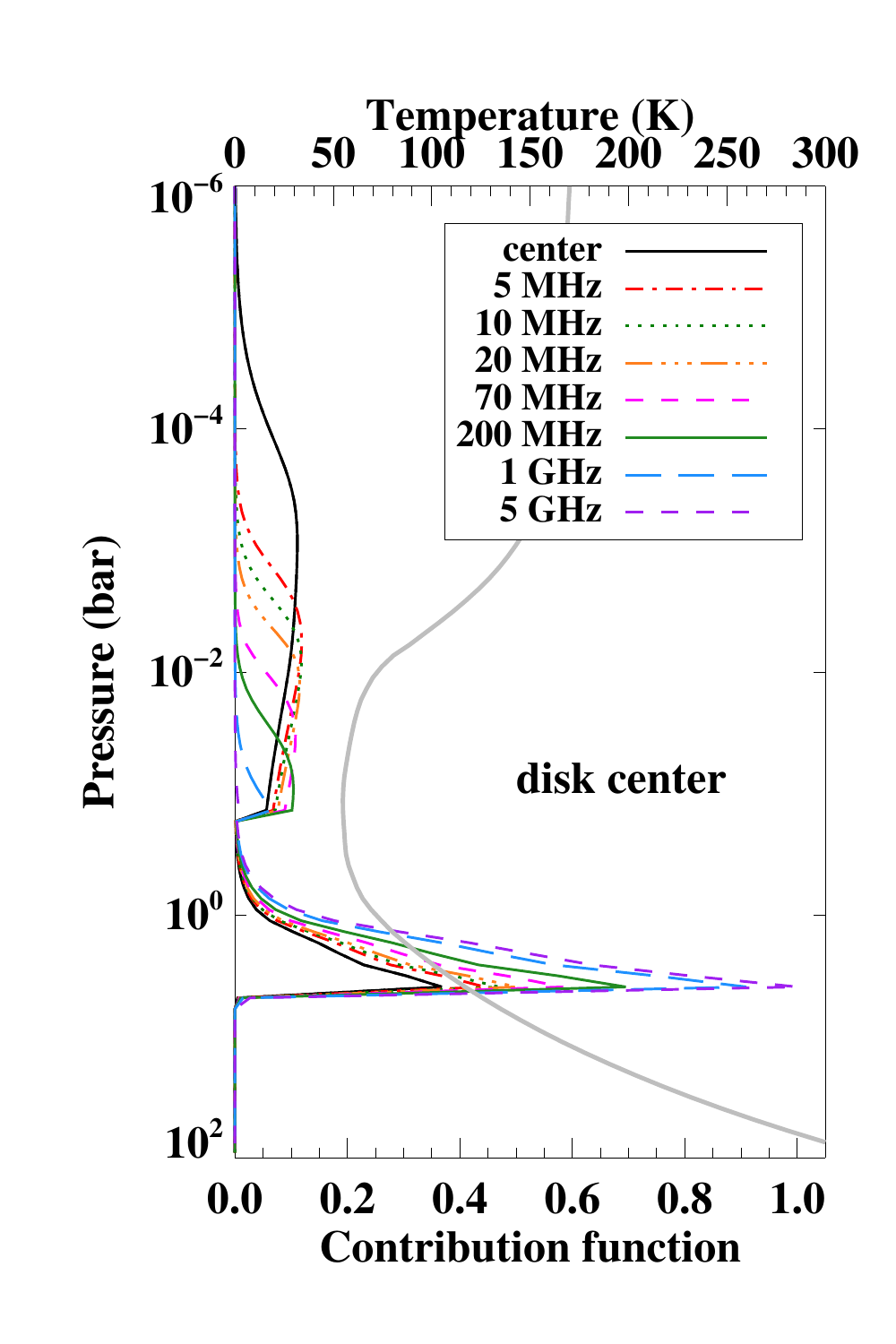}
\includegraphics[width=0.4\textwidth]{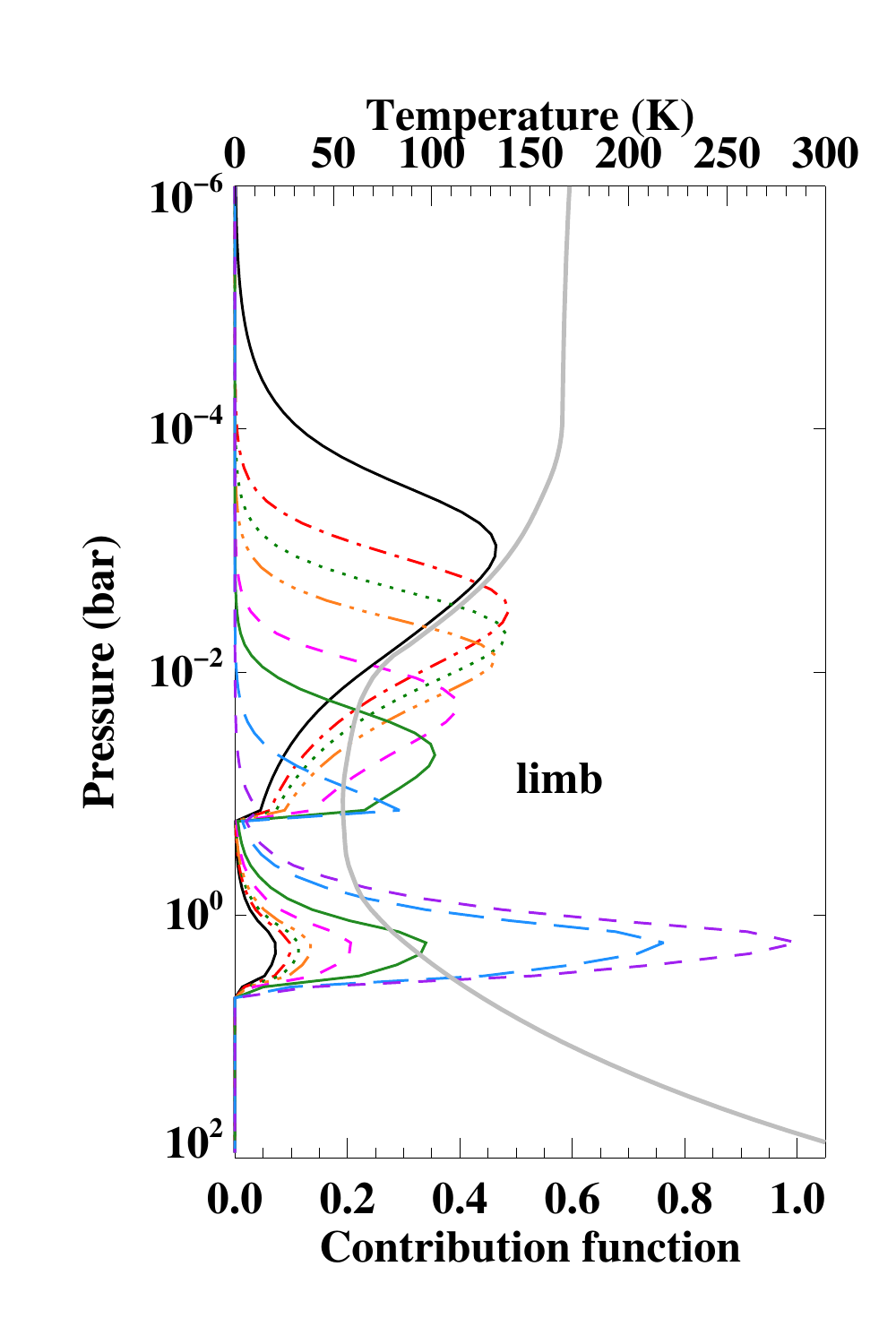}
$
\end{center}
\caption[Limb and disk center contribution functions]{\label{fig:mapcontrib}  Contribution functions for the CO (2--1) line, illustrating the altitudes contributing to the observed intensity for a selection of offsets from 0 to 5 GHz from line center. Contribution functions have been produced from the radiative transfer model, for disk center (selected to be where the viewing angle $\mu$ (the cosine of the emission angle) is greater than 0.9, shown left), and near the limb ($\mu<0.45$, right). The model was produced assuming a CO profile with 1.1 ppm CO in the stratosphere (at pressure less than 0.16 bar) and no CO in the troposphere. CO opacity is greatest near line center; and emission at this frequency comes from the highest altitudes in the atmosphere (black line).  The cutoff in the CO abundance is responsible for the sharp decrease in the contribution functions near the tropopause. Neptune's thermal profile has been plotted for reference (grey, see Section \ref{sec:comodel} for more information).}
\end{figure*}

To differentiate between these scenarios, models of Neptune's disk-integrated vertical CO profile have been produced using observations of the CO (1-0) \citep{luszcz13}, (2-1) \citep{lellouch05,luszcz13} and (3-2) \citep{hesman07} rotation lines at high spectral resolution (1.25-4 MHz) over a wide (8-20 GHz) frequency range.  As illustrated in Fig. \ref{fig:mapcontrib} for the CO (2-1) line at 230.538 GHz, such observations are necessary to characterize the full vertical CO profile, detecting emission from pressures below 0.1 mbar at line center, up to several bars in the far wings. From their respective studies, \cite{lellouch05} and \cite{hesman07} find substantial tropospheric CO abundances of $0.5\pm0.1$ and $0.6\pm0.4$ parts per million (ppm). The analysis of \cite{luszcz13}, which favors a warmer temperature profile than \cite{lellouch05} and \cite{hesman07}, produces a lower best-fit tropospheric CO abundance of $0.1^{+0.2}_{-0.1}$ ppm. As first described by \cite{prinn77}, the observed tropospheric CO mole fraction represents the equilibrium abundance at the CO `quench level', which is defined as the depth at which the timescale for vertical mixing is equal to the timescale for chemical conversion of CO into CH$_4$. The equilibrium CO mole fraction is directly proportional to the equilibrium abundance of H$_2$O, and under the conditions of Neptune's deep atmosphere, nearly all the gas phase oxygen is contained in water.  Therefore, the CO abundance of tropospheric CO acts as a probe of Neptune's global oxygen abundance. \cite{luszcz13} find that a tropospheric CO mole fraction of 0.1 ppm implies a global oxygen enrichment of at least 400, and likely more than 650 times the protosolar O/H value. Note, though, that the \cite{luszcz13} data are also consistent with 0.0 ppm of CO in the troposphere, in which case they do not constrain the global oxygen abundance.  In addition to the CO abundance measured in the troposphere,  \cite{lellouch05}, \cite{hesman07}, and \cite{luszcz13} all find that the CO line shape is best fit by a CO abundance profile that increases in the stratosphere, which suggests that  infall must also contribute to Neptune's observed CO abundance. Based on the atmospheric CO/H$_2$O ratio, \cite{lellouch05} proposed that a recent large cometary impact could be responsible for Neptune's observed stratospheric CO abundance; however, comets of the necessary size are expected to be exceedingly uncommon.  \cite{luszcz13} find that a constant influx of (sub)kilometer-sized comets could supply the observed stratospheric abundance of CO.

Spatially resolved maps of Neptune at centimeter wavelengths have been obtained by several authors  \citep{depater91,martin06, martin08, hofstadter08}. \cite{martin06,martin08} and \cite{hofstadter08}  find a substantial (tens of K) increase in the 1.3-2 cm brightness temperature near the south pole. Such a temperature enhancement would result if dry air subsides at this location, which would cause a decrease in the atmospheric gas opacity so that warmer, deeper layers of the planet are probed. This observation is therefore consistent with a global circulation pattern in which air rises at mid- southern and northern latitudes and subsides near the equator and south pole.  Recently, \cite{butler12} presented maps at 1 cm obtained with the upgraded VLA, with a resolution of better than 0.1'': they observe that Neptune's bright polar cap extends from the pole to 70$^\circ$S. They also see evidence for equatorial brightening, which would be consistent with the circulation pattern outlined above. 
 \begin{figure*}[t]
\begin{center}$
\includegraphics[width=0.4\textwidth]{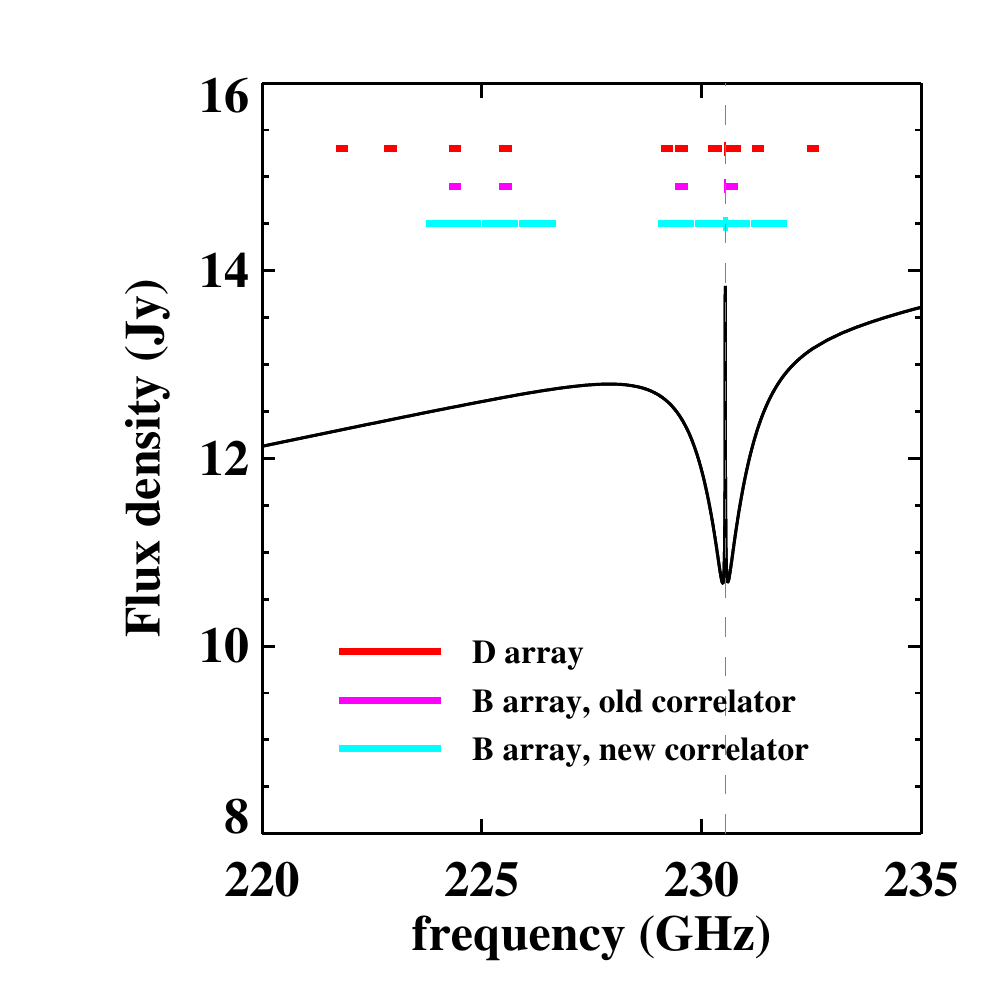}
\includegraphics[width=0.4\textwidth]{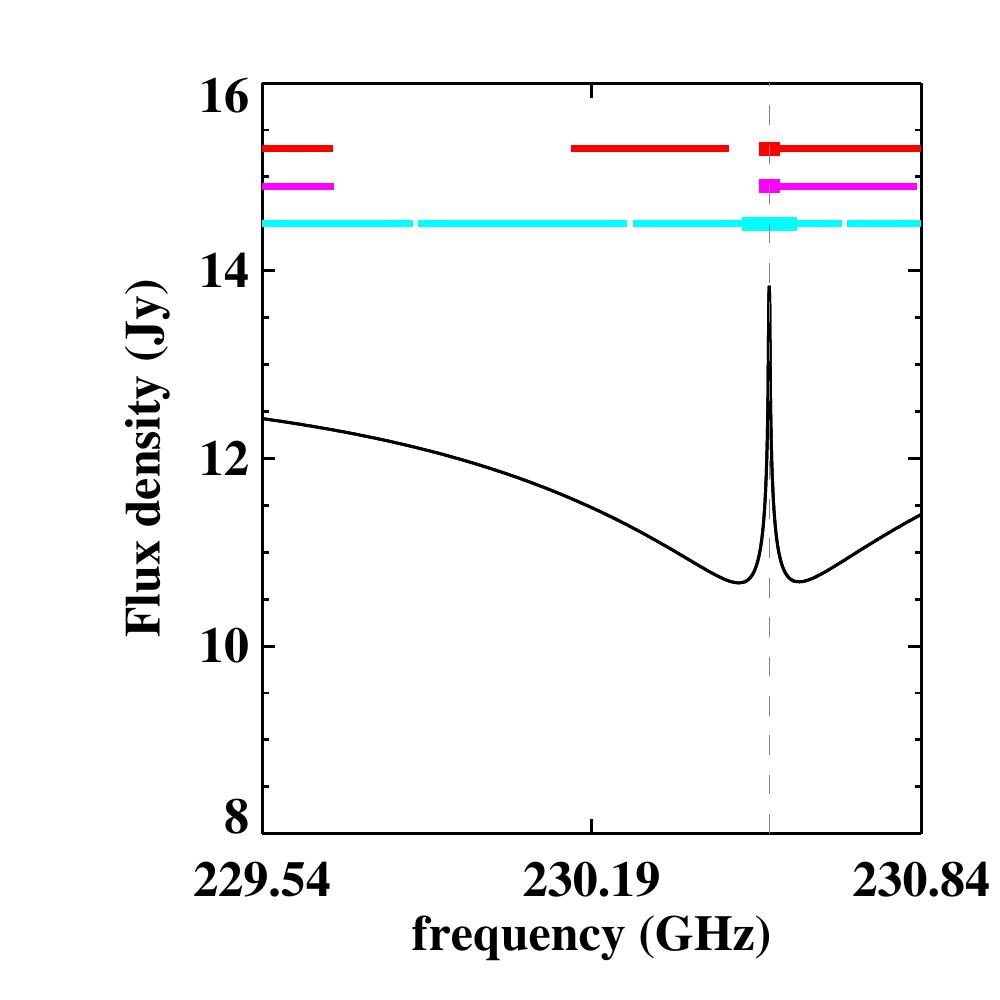}
$
\end{center}
\caption[Spectral coverage]{\label{fig:coverage} Spectral coverage of our observations; for reference the model disk-integrated spectrum is shown in black. The short baseline D-array data windows are indicated in red. The long baseline B-array bands are shown in pink (prior to 2011) and cyan (2011). For each correlator setup, the narrow band is shown as a thicker line than the wide bands. As described in Section \ref{sec:obs}, our analysis focuses on the frequencies where we have the most data, which are those spanned by the B-array data prior to 2011 (pink). The plot on the right is zoomed in on line center.}
\end{figure*}
In this paper, we present the first spatially-resolved measurements of Neptune at millimeter wavelengths,  originally reported by \cite{luszcz10}. Our data set spans a range of frequencies, from the center of the CO (2-1) line at 230.538 GHz (1.3 mm) to a maximum offset of 6 GHz from line center, where continuum emission is detected.  The motivation behind these observations was to look for latitudinal variations in the CO abundance, which would provide additional information about the pattern of CO infall/production. This was observed in the case of Jupiter after the impact of comet Shoemaker-Levy 9 (SL9); \cite{moreno03} estimated that for SL9, latitudinal variations in the CO abundance would persist for roughly a decade. Furthermore, the CO abundance at a given latitude depends on the rate of vertical and meridional mixing, and could therefore act as a tracer of the large-scale circulation. The spectrum in the CO line is also affected by the temperature profile; therefore maps in the CO line will be affected by horizontal variations in temperature. In the 1.3 mm continuum, the intensity depends on the gas opacity at depths of 1.1-4.7 bar which may vary with latitude due to the large-scale circulation pattern.

\section{Observations}\label{sec:obs}

We observed Neptune with the Combined Array for Research in Millimeter-wave Astronomy (CARMA), located in the Inyo Mountains of eastern California \footnote{CARMA is located at Cedar Flat, CA, at elevation 2200 m; latitude 37.3¡; longitude -118.1¡}. CARMA consists of eight 3.5-meter antennas, nine 6-meter antennas and six 10-meter antennas; our observations were performed with the 6- and 10-meter antennas only, for a total of 105 baselines and 725 m$^2$ of collecting area. CARMA can be configured in 5 standard patterns `A--E', with the `A' configuration being the most extended and therefore producing the highest angular resolution; and the `E' configuration being the most compact (lowest resolution). In order to spatially resolve Neptune's 2.2'' disk yet still `see' the object and have enough signal per synthesized beam, we observed in the `B' configuration, which has baselines of 82-946 meters and a synthesized beam of approximately 0.35" at 230 GHz. Projected antenna separations ranged from 30 to 670 k$\lambda$. 
\begin{table*}[htb!]
\scriptsize
\begin{center}
\setlength{\tabcolsep}{0.05in}

\caption{}{B-array correlator setup, prior to 2011. This setup corresponds to configuration `d' as specified in Table 1 of \cite{luszcz13}. In this paper only data within the frequencies spanned by this correlator setup are considered. \label{tab:config}}

  \vspace{\baselineskip}

 \footnotesize
\begin{tabularx}{1.\textwidth}{l l l l l l}
\hline
Band center frequency & Nominal bandwidth\footnote{CARMA correlator bands labeled as having a nominal width of 500 MHz bands actually have a total width of 468.75 MHz} & Final bandwidth\footnote{after flagging of edge channels} & Start frequency$^{\text{b,}}$\footnote{frequency of band edge, NOT frequency at the center of the edge channel}&End frequency$^\text{b,c}$& Channel spacing \\
224.387 GHz & 500 MHz	& 281 MHz	& 224.247 GHz	&	224.528 GHz	&	31.25 MHz\\
225.537 GHz & 500 MHz 	& 281 MHz	& 225.397 GHz 	&	225.678 GHz	&	31.25 MHz\\
229.538 GHz & 500 MHz	& 281 MHz	& 229.397	GHz	&	229.679 GHz	&	31.25 MHz\\
230.688 GHz & 500 MHz	& 281 MHz	& 230.547 GHz 	&	230.829 GHz	&	31.25 MHz\\
230.538 GHz & 62 MHz 	& 42.0 MHz	& 230.517 GHz 	&	230.559 GHz	& 0.9766 MHz \\
\hline
\end{tabularx}
\end{center}
\end{table*}

Observations were carried out over a total of 13 days: one test observing block, or `track', was taken in December 2008, followed by eight tracks in the winter of 2009-2010 and four in January 2011. Of these, two of the datasets from 2011  were of poor quality and were not included in the analysis. Each dataset consists of a 15-minute observation of the bright quasar 3C454.3 for passband calibration, followed by a series of observing cycles of 8 minutes on Neptune and 2 minutes (prior to  2011) or 3 minutes (in 2011) on a nearby quasar (2229-085 in December 2008 and January 2011; 3C446 in winter 2009-2010) to be used for calibration of the atmospheric and instrumental gains \footnote{phase calibrators were typically located at a separation of 10--15 deg from Neptune. For more information on these calibrators, see http://carma.astro.umd.edu/cgi-bin/calfind.cgi}. The weather conditions during the observations were generally fair, with root mean square (rms) path errors ranging from 100 to 325 $\mu$m on a 100-m baseline, and an average zenith optical depth of 0.18. The total time on source in the B-array configuration was 28 hours.  To provide information at shorter baselines, we combined our B-array data with 15 hours on source in the more compact D-array configuration from spring 2009; these data have projected antenna separations of 5--84 k$\lambda$ and are described in \cite{luszcz13}. 

Prior to upgrades in 2010, the CARMA correlator offered three dual bands, or spectral windows, with configurable width of 500, 62, 31, 8 or 2 MHz. Each band could be placed independently anywhere within the 4 GHz IF bandwidth, and appears symmetrically in the upper and lower sidebands of the first local oscillator. For our B-array observations, we configured one correlator band to 62-MHz bandwidth and centered on the CO (2--1) line at 230.538 GHz; the channel spacing in this `narrow' band was 0.9766 MHz. The remaining two bands were configured to maximum (500 MHz) bandwidth and placed at offsets from line center; the channel spacing in each of these `wide' bands was 31.25 MHz. The configuration of the correlator during our 2008-2010 B-array observations is described in Table \ref{tab:config}. Due to upgrades to the correlator in late 2010, the channel spacing decreased in both the 62-MHz and 500-MHz bands, and five additional dual bands were available during our 2011 observations. These additional bands were all configured to maximum bandwidth mode, allowing for more continuous coverage of the wide CO line. Figure \ref{fig:coverage} illustrates the location of the correlator bands for the B-array observations prior to and after the correlator upgrade, as well as the frequency coverage of the D-array data. In this paper, we restrict our analysis to the frequencies spanned by the original correlator setup (Table \ref{tab:config}), where we have the most data and therefore the greatest sensitivity.

We perform editing and calibration on each raw visibility dataset using the Multichannel Image Reconstruction, Image Analysis and Display \citep[MIRIAD;][]{sault11} software package. After flagging edge channels and poor quality data, we perform passband calibration. Then we correct for atmospheric and instrumental effects on the observed visibilities. We determine the time-dependent antenna-based gains by self-calibrating the wide-band data and then applying the gain solutions to the full dataset. The absolute flux scale of each dataset is set by scaling the antenna gains for each day of B-array data so that the amplitudes of the visibilities match the amplitudes of the D-array data at overlapping (u,v) distances\footnote{The u and v coordinates describe the East-West and North-South components of the projected interferometer baseline. The (u,v) distance is defined as the projected baseline length.}. A complete description of the data reduction is provided in Appendix \ref{sec:reduction}.

\section{Model}\label{sec:comodel}
A series of model image cubes were created using the line-by-line radiative transfer code described in \cite{luszcz13}, and using Neptune's atmospheric properties as described therein. We adopt the thermal profile of \cite{fletcher10} in the upper atmosphere, and an H$_2$S abundance corresponding to 50 times the protosolar S/H abundance.  As our nominal case, we model the CO (2-1) absorption assuming that the CO abundance is vertically distributed according to the best-fit solution of \cite{luszcz13} for the above-mentioned atmosphere \footnote{Several of the atmospheric models in \cite{luszcz13} do not include any H$_2$S; here, we adopt the best-fit solution for the model which has a 50 times solar abundance of H$_2$S.}: 1.1 ppm of CO at altitudes above 0.16 bar and 0.0 ppm of CO deeper than this level. At each of 155,575 locations on the disk (0.005'' pixels), we integrate the equation of radiative transfer for the appropriate viewing angle $\mu$ (the cosine of the emission angle), accounting for the Doppler shift due to the planet's rotation \citep{moreno01}. These high-resolution models can be converted from brightness temperature units into Jy/pixel, to be used as starting models for the deconvolution process (Section \ref{sec:imaging}, Appendix \ref{app:imagedecon}). They can also be rebinned to the coarser map resolution and convolved with a Gaussian beam, and directly compared with the data (Section \ref{sec:coresult}). For the latter purpose, alternative models are also produced by assuming the same vertical CO structure (the same transition pressure, and no tropospheric CO), but varying the stratospheric CO abundance; or alternatively by maintaining the nominal CO structure and abundance but varying the temperature or composition (other than CO) with latitude. 

\section{Imaging}\label{sec:imaging}
\begin{table}[b]
\scriptsize
\begin{center}
\setlength{\tabcolsep}{0.05in}

\caption{}{Characteristics of final maps. Noise in each final map is calculated as the root mean square of the noise in the input maps, divided by the square root of the number of input images. See Appendix \ref{app:error} for a discussion of noise estimation in the original (imaging) maps. \label{tab:mapfreq}}
 
 \vspace{\baselineskip}
 \footnotesize
\begin{tabularx}{.5\textwidth}{p{2.8cm} p{2.7cm} p{2.4cm}}
\hline
Average frequency offset from line center & Effective frequency width\footnote{Total frequency width averaged to create the map.} (MHz)& Noise in averaged map (mJy/beam)\\
0.0 MHz  	&4.6  	&28.3 \\
4.6 MHz  	&9.2 	&19.9\\
9.2 MHz  	&9.2 	&20.2\\
13.8 MHz 	&9.2	&19.7\\
18.4 MHz 	&9.2	&19.3\\
66 MHz		&125	&8.13\\
213 MHz	&125	&7.62\\
1.00 GHz	&250	&6.01\\
5.58 GHz	&500	&3.71\\
\hline
\end{tabularx}
\end{center}
\end{table}

After combining all tracks of B- and D-array visibility data, we produce images of Neptune in four steps: first we use our visibility data to make a set of maps of the estimated sky intensity distribution. The FWHM (full width at half maximum) of the image point spread function, or synthesized beam, is in the range of 0.33--0.37'' x 0.37--0.39'', depending on frequency; the pixel size used in the maps is 0.09''.  Next, we estimate the noise in our maps due to calibration errors and atmospheric fluctuations. We find that the noise level is approximately three times greater than one would expect from thermal noise from the atmosphere and receiver system, likely because of the unfavorable weather during observations. We then perform deconvolution to remove the response to source structure in the side lobes of the synthesized beam. Finally, we average deconvolved images that are at similar frequency offsets from the CO (2-1) line center, in order to increase signal-to-noise without compromising our sensitivity to the vertical structure in Neptune's atmosphere, i.e., we combine channels which have similar contribution functions (Fig. \ref{fig:mapcontrib}). We also bin the data according to latitude to look for meaningful latitudinal trends in Neptune's observed brightness. Table \ref{tab:mapfreq} lists the central frequency and total frequency width of each final image, along with our estimate of the noise in units of mJy/beam. The details of the imaging process are presented in Appendix \ref{app:imagedecon}; this description includes a discussion of the final error estimates and the binning technique. A comparison of deconvolution methods is given in Appendix \ref{app:decon}. 

\begin{figure}[htb!]
\begin{center}$
\includegraphics[width=0.3\textwidth]{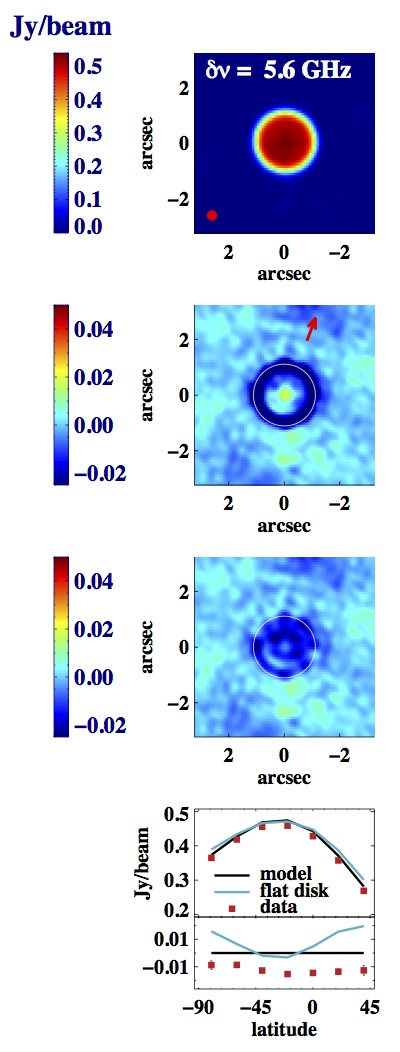}
$
\end{center}
\caption[Average continuum map]{\label{fig:avgcont}  Average of the four maps with frequency offsets from line center of 5--6 GHz. Emission at these frequencies is primarily millimeter continuum. Top image is the final, deconvolved and averaged map. The average beam is indicated by the red circle in the bottom left corner of the image. Second image from the top shows the same data with a (beam-convolved) uniform brightness disk subtracted. The white circle indicates the location of the planet's limb, and the red arrow indicates the direction of the rotation axis. The third image from the top has the beam-convolved nominal (horizontally uniform composition and temperature) model subtracted. In the plot (bottom) we zonally average the image (red points) and nominal model (black) as described in Section \ref{sec:comodel}. For reference, we also show the average of the beam-convolved uniform brightness disk as a function of latitude (blue). The differences between the nominal model and the data (and between the nominal model and a uniform brightness disk) are shown in the bottom part of the plot.}
\end{figure}

\begin{figure*}[htb!]
\begin{center}$
\includegraphics[width=0.65\textwidth]{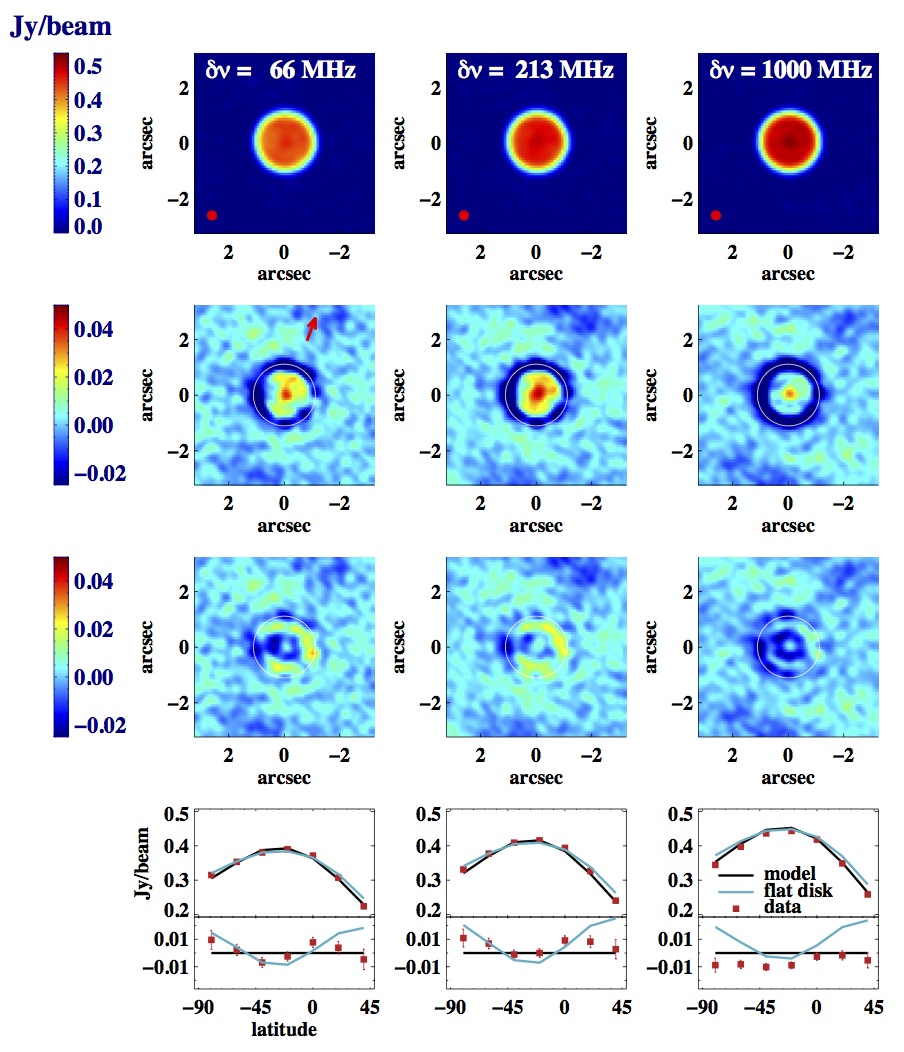}
$
\end{center}
\caption{\label{fig:avgwide} Same as Fig. \ref{fig:avgcont}, except for $\delta \nu=$ 66, 213, and 1000 MHz.    }
\end{figure*}

\begin{figure*}[htb!]
\begin{center}$
\includegraphics[width=1.0\textwidth]{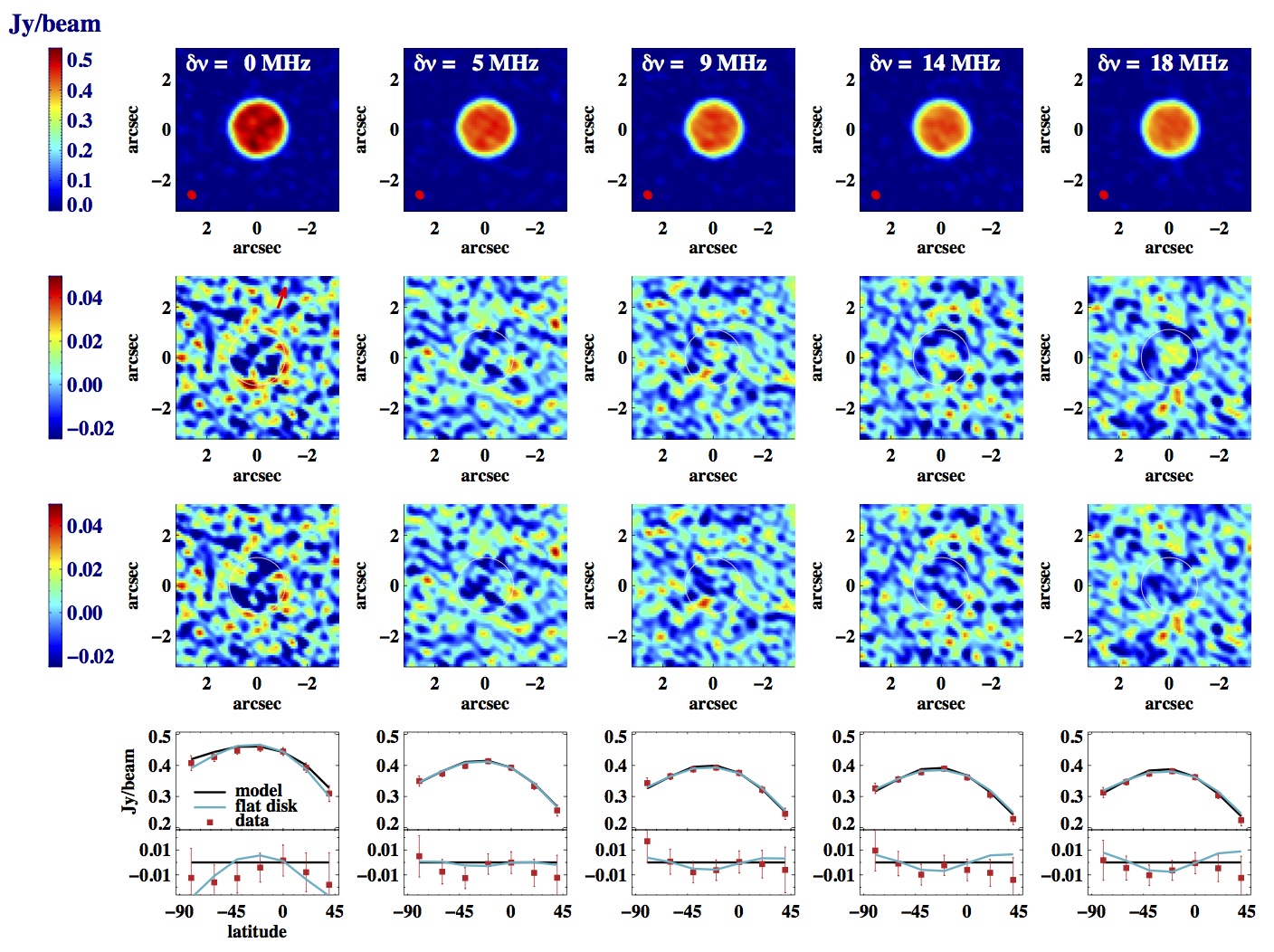}
$
\end{center}
\caption{\label{fig:avgnar} Same as  Fig. \ref{fig:avgcont}, except for $\delta \nu =$ 0--18 MHz. These images average over the smallest frequency interval (4.6 MHz for $\delta \nu =0$, 9.2 MHz for the others), and therefore have the highest level of noise, as indicated by the error bars in the plots. }
\end{figure*}

The images and latitude-binned plots are presented in Figs. \ref{fig:avgcont}--\ref{fig:avgnar}. The top panel within each column is the final, averaged map at the indicated frequency offset ($\delta \nu$) from line center. The FWHM of the center of the synthesized beam, which denotes the spatial resolution of the map, is indicated by the filled red ellipse in the lower left corner. The arrow points in the direction of Neptune's north pole. Below each map, we also present two `residual maps', which illustrate the variations between the data and a `flat' (uniform brightness) disk; and from the nominal, horizontally uniform model described in Section \ref{sec:comodel} (second and third from the top, respectively). The first of these residual maps emphasizes the variations across the disk due to changes in emission angle (limb brightening/darkening); the latter highlights horizontal variations in the composition and/or thermal profile. Below each set of images, we present the latitude-binned data, as compared to the latitude-binned flat disk and model. To look for spatial variations in the observed atmospheric properties, we also show the residuals between the horizontally uniform nominal model and the data. These figures are discussed further in the following section.

\section{Results}\label{sec:coresult}
\subsection{Continuum variations}\label{sec:continuum}
The image at an average offset of 5.6 GHz from line center (Fig. \ref{fig:avgcont}) is representative of the millimeter continuum. The most noticeable spatial variation we observe in the data is significant limb-darkening, as is predicted by the radiative transfer mode. We also observe artifacts induced by calibration errors and atmospheric fluctuations, which create a noticeable `ringing' pattern in the images. Averaging the data with respect to latitude mitigates this issue, but higher-quality observations are necessary to produce images without these artifacts.

 \begin{figure*}[htb!]
\begin{center}$
\includegraphics[width=0.6\textwidth]{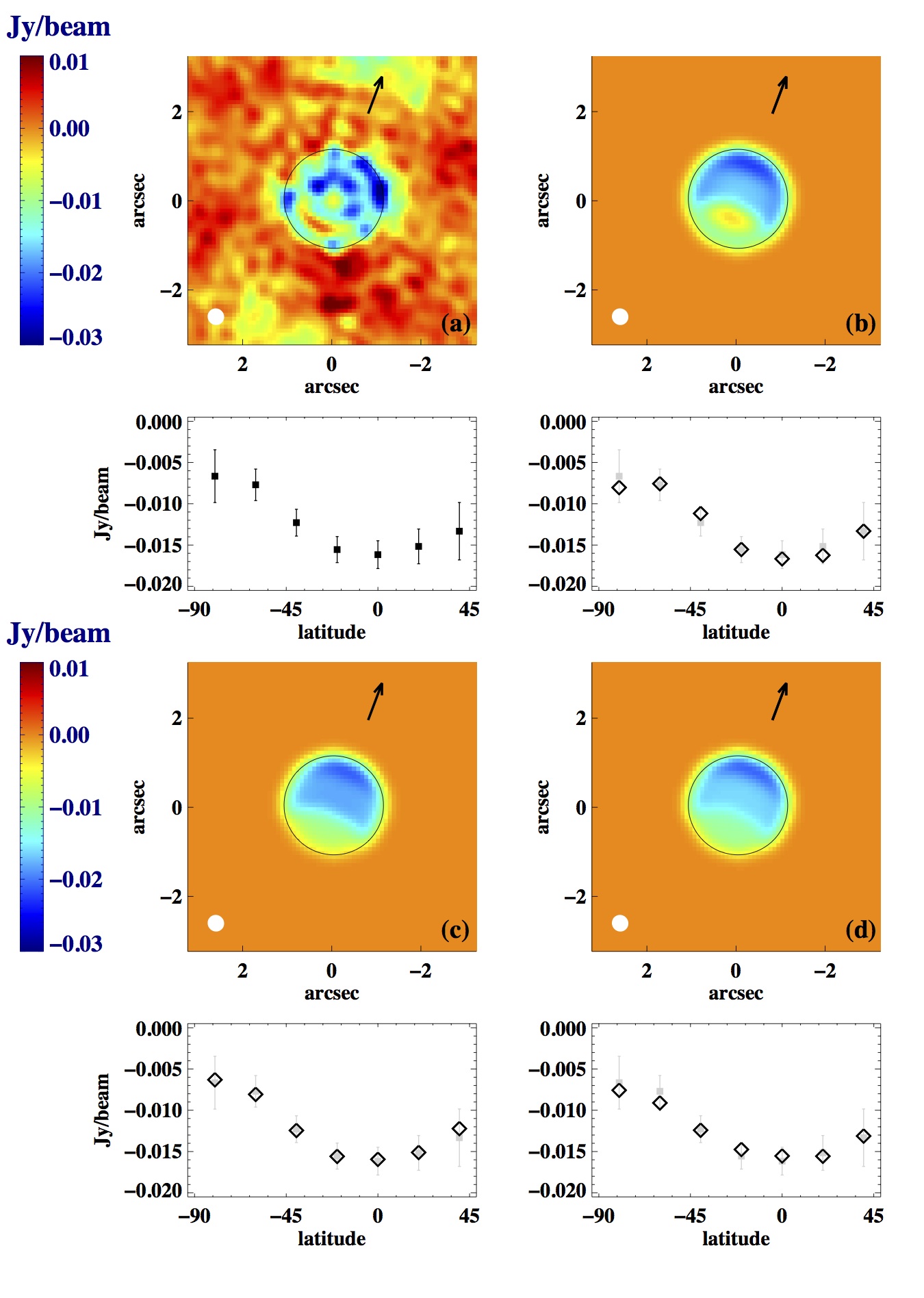}
$
\end{center}
\caption{\label{fig:continuum} \footnotesize Latitude variations in the continuum map. Panel (a) is adapted from Fig. \ref{fig:avgcont}, with a different scaling of the color bar and $y$ plot axis. The image is the $\delta \nu =$ 5.6 GHz data with the beam-convolved, horizontally uniform composition and temperature model subtracted. The plot shows the latitude-binned residuals of the data$-$model.  The filled white circle in the bottom left corner represents the full width at half maximum of the beam. The black circle indicates the location of Neptune's limb, and the black arrow indicates the direction of the rotation axis. Panel (b) shows a (scaled) model in which the H$_2$S opacity has the nominal value from 10$^\circ$N to 40$^\circ$N, 0.3 times the nominal value from 40$^\circ$S to 10$^\circ$N, and 0.03 times the nominal value between 90$^\circ$S and 40$^\circ$S. The nominal model has been subtracted as in (a). The plot beneath the image shows the same data, binned in latitude (black diamonds). The data points from (a) are in grey. Panels (c) and (d) are the same as (b), with different  models: (c) shows a model where the tropospheric CH$_4$ mole fraction is 0.044 (twice the nominal value of 0.022) from 10$^\circ$N to 40$^\circ$N,  .022  from 40$^\circ$S to 10$^\circ$N, and 0.0055 (25\% of nominal) between 90$^\circ$S and 40$^\circ$S, is shown; (d) shows a (scaled) model where the equilibrium hydrogen fraction is 0.6 from 10$^\circ$N to 40$^\circ$N,  0.8 from 40$^\circ$S to 10$^\circ$N, and 1.0 between 90$^\circ$S and 40$^\circ$S. }
\end{figure*}

Additionally, the data in Fig. \ref{fig:avgcont} show two trends in latitude with respect to the nominal (horizontally constant composition and temperature) model. The first is an overall offset in the brightness between the continuum data and nominal model. We compare the integrated flux density of the data and model and find that the difference is  $-0.41 \pm 0.02$ Jy, or $-2.9 \pm 0.1$ K. This 3\% continuum offset could be due to issues in calibration as well as real deficiencies in our model of the continuum; this is discussed more thoroughly in \cite{luszcz13}. Here we concentrate on the second observed trend in the data: latitudinal variations in the continuum brightness. We observe that Neptune's south pole appears to be brighter than mid- and northern latitudes; this is highlighted in panel (a) of Fig. \ref{fig:continuum}, which is a scaled version of Fig. \ref{fig:avgcont}. To determine the significance of this result, we calculate $\chi^2$:
\begin{equation}
\chi^2= \sum^{M-1}_{m=0} \frac{\delta y_m^2}{\sigma_{meas,m}^2}
\end{equation}
where $M$ is the number of data points in the fit, the parameter $\delta y_m$ is the difference between the data and model values for point $m$, and $\sigma_{meas,m}$ is the measurement error for point $m$. We use the latitude-binned data for this calculation, so that $\delta y_m$ is the difference between the average flux density of the data and the model in bin $m$ (in Jy/beam); and $\sigma_{meas,m}$ is 3.71 mJy/beam (the noise estimate in the map within a single beam), divided by the square root of the number of beams in bin $m$, as shown by the error bars in Fig. \ref{fig:continuum}. Since points from neighboring latitude bins are correlated (the width of each bin is roughly equal to one beam), we include only every other point in the statistical fit, starting with the data point nearest the south pole, such that $M=4$. The model that we use in calculating $\chi^2$ is the nominal model scaled by a constant factor of 0.97 to best match the data. We then compare the value of $\chi^2$ to the probability $p$ of measuring that value for a set of experiments with a degree of freedom
\begin{equation}
DOF= M-N
\end{equation}
where $N$ is the number fit parameters (in this case, $N=1$). We find that for these data compared to the scaled, horizontally uniform nominal model, the calculated $\chi^2= 8.8$.  For a fit of 3 degrees of freedom there is a probability $p= 0.03$ of a $\chi^2$ value this high occurring due to chance. We therefore conclude that real spatial variations in Neptune's  millimeter continuum are likely, and we explore the mechanisms that could cause such variations.

The contribution function at an offset of 5 GHz from the CO line center peaks at a depth of 4 bar, with most of the emission originating from depths of 1.1-4.7 bar. We expect that variations at these frequencies are likely due to  opacity variations, as opposed to horizontal temperature variations, as the temperature profile is likely adiabatic in the troposphere. A decrease in gas opacity near the south pole would mean that warmer, deeper layers are probed. To estimate the magnitude of change in optical depth that is required produce the observed flux densities, we first match the data by scaling the total optical depth in the model at the relevant pressures. We find that decrease of about  30\% in the south at pressures greater than 1 bar is sufficient to match the observed latitudinal gradient. If we vary the opacity only at pressures greater than 2 bar, we require the opacity to be lower by a factor of 2 in the south (e.g. from  the nominal opacity at 40$^\circ$N to 0.5 times nominal near the south pole). If we restrict the opacity variations to  pressures greater than 3 bar, we require a factor of 5-10 decrease in the opacity in the south and if variations are only at pressures greater than 4 bar, the opacity must be of order 50 times lower in the south than at 40$^\circ$N to match the observations. Therefore, the total opacity change across the disk implied by our data depends on the pressures at which the opacity changes, which is determined by what opacity source is responsible. We investigate how variations in three atmospheric properties might cause the observed trend: the H$_2$S abundance, the CH$_4$ abundance and the  H$_2$ {\it ortho/para}  ratio. 

The dominant opacity sources in our model at pressures of 1.1-4.7 bar are H$_2$S opacity and H$_2$ collision-induced absorption (CIA). For an H$_2$S abundance of 50 times solar, H$_2$S opacity begins to dominate over CIA at pressures greater than 3.3 bar. The precise value and variation of the H$_2$S abundance in Neptune's troposphere is poorly constrained, though cm-wavelength observations suggest it may vary with latitude according to the meridional circulation pattern \citep{martin08,hofstadter08,butler12}. We find that we can reproduce the latitudinal behavior observed in our data by varying the H$_2$S opacity  by a factor of $\sim$30 from the south pole to 40$^\circ$N. Figure \ref{fig:continuum}b shows the expected residuals for a case where the H$_2$S opacity has the nominal value from 10$^\circ$N to 40$^\circ$N, 0.3 times the nominal value from 40$^\circ$S to 10$^\circ$N, and 0.03 times the nominal value between 90$^\circ$S and 40$^\circ$S. This model has also been scaled by a factor of 0.95 to match the integrated flux of the data. 

A decrease or increase in CH$_4$ can also affect the brightness temperature, even though methane on its own does not contribute directly to the opacity \citep{depater93}. CH$_4$ condensation can change the adiabatic profile; however, we adopt a dry adiabat in the upper atmosphere \citep{luszcz13}, so CH$_4$ does not affect our models in this way. Although the CH$_4$ mole fraction in the troposphere is only 2.2\% \citep{baines95}, the collision-induced absorption coefficient for H$_2$-CH$_4$ pairs is roughly 20 times higher than for H$_2$-H$_2$ pairs under the relevant conditions. As a result,  H$_2$-CH$_4$ CIA accounts for as much as 35\% of the total optical depth in the 1-4 bar region. As for H$_2$S, we determine the magnitude of variations that are required to reproduce the observed intensity gradient: we find that  the south polar region must be depleted in tropospheric CH$_4$ by a factor of order 10 relative to 40$^\circ$N. Figure \ref{fig:continuum}c shows the expected residuals for a case where the tropospheric CH$_4$ mole fraction is 0.044 (twice the nominal value of 0.022) from 10$^\circ$N to 40$^\circ$N, 0.022 from 40$^\circ$S to 10$^\circ$N, and 0.0055 (25\% of nominal) between 90$^\circ$S and 40$^\circ$S. The disk-averaged CH$_4$ mole fraction for this example is 0.021, which is close to the expected nominal value. As in Fig. \ref{fig:continuum}b, a scale factor (in this case, 0.97) has been applied to the model to match the data.

Finally, we consider variations in the H$_2$ {\it ortho/para} ratio as a source of continuum intensity variations. As in \cite{luszcz13}, our nominal model assumes `intermediate' hydrogen as the nominal case. That is, the {\it ortho} and {\it para} states of hydrogen are in equilibrium at the local temperature, but  the specific heat is near that of `normal' hydrogen. This situation is described in \cite{massie82}.  Fast vertical mixing from the deep interior, however, could bring the {\it ortho/para} ratio closer to the 3:1 ratio expected for normal hydrogen. We investigate the effect of variations in the fraction of {\it para}-H$_2$ on the opacity, while assuming the adiabatic profile for normal hydrogen everywhere. We find that normal hydrogen has a higher opacity than `equilibrium' hydrogen. The average intensity is 5-6 K lower when we assume normal hydrogen rather than equilibrium hydrogen. This implies that variations in the  {\it ortho/para}  ratio are more than capable of causing intensity variations of the magnitude observed in our data. We can match the observed latitudinal variations by decreasing the fraction of hydrogen in the equilibrium state from 1.0 at the south pole to 0.6 at 40$^\circ$N (Fig. \ref{fig:continuum}d), and scaling this model by a factor of 0.98.

\subsection{Variations near the tropopause}\label{sec:covar}

\begin{figure*}[htb!]
\begin{center}$
\includegraphics[width=0.7\textwidth]{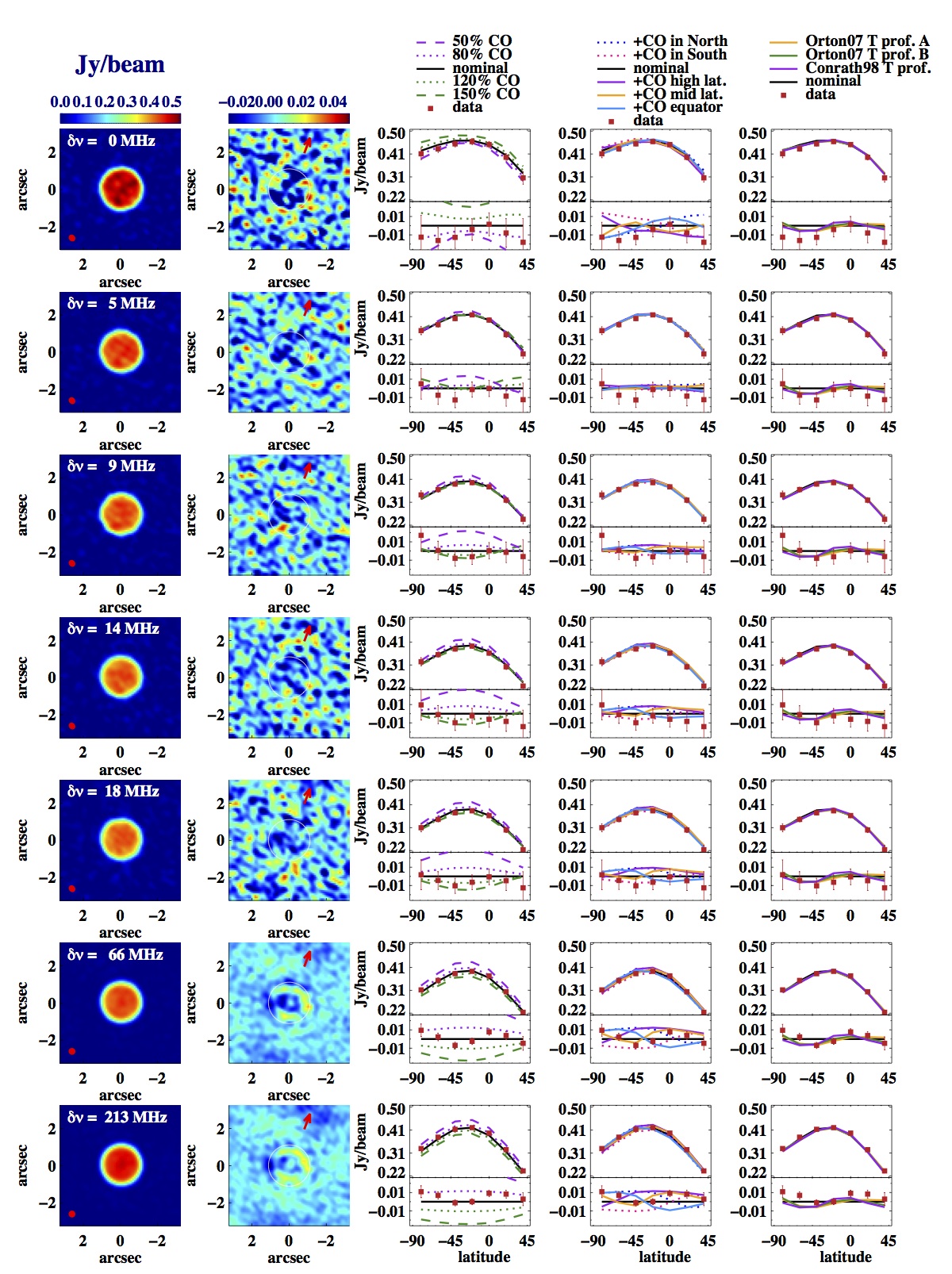}
$
\end{center}
\caption[Comparison of data and models, offsets of 0-213 MHz]{\label{fig:comparison}  \footnotesize Comparison of the zonally-averaged data and models, for frequency offsets of 0--213 MHz. First two columns are the \CLEAN\ maps and nominal-model subtracted maps. The full width at half maximum of the beam is shown by a filled red circle. In the difference maps, the limb is indicated by a white circle and the direction of the rotation axis is shown by a red arrow.  For each set of data$-$model comparisons, the top portion of each plot shows the binned data (red squares) with errors and several comparison models (identified at the top of each column). Below each plot is a difference plot, showing the same data and models with the nominal model subtracted, highlighting spatial variations in brightness due to variations in opacity and/or temperature. The first column of plots shows a selection of horizontally uniform CO, uniform temperature models (Section \ref{sec:counif}). The second column of plots presents horizontally uniform temperature, meridionally varying CO models (Section \ref{sec:covarco}). The final column presents three horizontally uniform CO, meridionally varying temperature models (Section \ref{sec:covartp}). }

\end{figure*}

The images at frequency offsets of 0--1000 MHz sense lower pressures than the continuum map, due to a contribution from CO to the opacity. Therefore these maps include information on altitudes near and above the tropopause at $\sim$0.1 bar (Fig. \ref{fig:mapcontrib}). Since observations towards the limb sense higher altitudes, we expect maps further from line center to be limb-darkened, whereas at line center, where the emission originates in the stratosphere, we expect limb-brightening. This is indicated in Figs. \ref{fig:avgwide} and \ref{fig:avgnar}, which show the data relative to the nominal radiative transfer model and to a uniform brightness (`flat') disk. We find that the limb-darkening can be observed above the noise in maps at offsets of 14-1000 MHz; the limb-brightening signal is marginally detected in the middle map (flat disk subtracted) at line center ($\delta \nu =$ 0).  Latitudinal trends in the data are more difficult to observe in these maps than in the continuum map, since these data are averaged over smaller frequency windows and therefore have lower signal-to-noise. This is particularly true for the images in Fig. \ref{fig:avgnar} which are typically averaged over only 9.2 MHz. We concentrate our analysis on the latitude-binned data at offsets of 0-213 MHz from line center, as shown in Fig. \ref{fig:comparison}. Interpretation of the 1000-MHz data is complicated by substantial contributions from continuum opacity sources; therefore these data are excluded from further analysis.

We note that the plots in Fig. \ref{fig:comparison} show a consistent latitudinal behavior: the south pole appears bright relative to our nominal model, southern mid-latitudes are relatively dark, and additional brightening occurs near the equator before the intensity falls off again at northern mid-latitudes. Variations at these frequencies could be due to variations in the CO abundance, variations in the temperature, or a combination of the two. We therefore compare these data to three sets of models with the following properties: horizontally uniform CO  and temperature profiles; a meridionally varying CO profile and horizontally uniform temperature profile; and a horizontally uniform CO profile and meridionally varying temperature profile. For each model of interest, we calculate $\chi^2$ as in Section \ref{sec:continuum}, using every other data point to avoid correlations between data. We determine $\chi^2$ for each frequency separately ($DOF=4$), as well as for the data from all seven mapping frequencies simultaneously ($DOF=28$). The results are presented in Table \ref{tab:chisq} and discussed below. For a fit with 4 degrees of freedom, a single measured value of $\chi_4^2$ has a probability $p= 0.05$ of being greater than 9.5. For a fit with 28 degrees of freedom, $\chi_{28}^2> 41.3$ with a probability $p=0.05$. Therefore, we reject fits with $\chi^2_4> 9.5$ and $\chi_{28}^2 > 41.3$.

\subsubsection{Horizontally uniform CO and temperature}\label{sec:counif}

We find that for the maps 0-18 MHz from line center, the nominal model, which maintains a uniform abundance of 1.1 ppm of CO in the stratosphere, is consistent with the data. At larger offsets from line center (66 and 213 MHz) the nominal model is rejected; at these wavelengths the total frequency width averaged to produce the images is greater, resulting in smaller error bars;  therefore the deviations are more statistically significant. When we consider the full dataset, we find that the nominal model has an acceptable value of  $\chi^2$ of 32.6. 

In addition to our nominal model, we produce models with 0.5, 0.8, 1.2 and 1.5 times the nominal CO abundance, with no latitudinal variation. Again we find that at large offsets from line center these models are inconsistent with the data. Figure \ref{fig:comparison} shows that, for these frequencies, and increase or decrease of more than 20\% in the CO abundance produces models that are too high/low to match the data at $\delta \nu=$ 66 and 213 MHz, for the given spatial resolution.  At smaller offsets, the effect of changing the CO abundance becomes more complex. At an offset of 5 MHz, for example, an increase or decrease in the CO abundance will lead to an overall brighter model: increasing the CO abundance would increase the limb brightening, while decreasing the CO abundance would increase the intensity at disk center. We conclude that models of this form (with no latitudinal variations in the CO abundance or temperature) are not capable of reproducing the observed trend in the CO abundance with latitude, and models with higher and lower CO abundances significantly decrease the goodness of fit of the full dataset.

\subsubsection{Meridionally varying CO, horizontally uniform temperature}\label{sec:covarco}
An interesting possibility is that the observed variations in Neptune's brightness could be due to latitudinal variations in the CO abundance, related to localized infall or circulation effects. We compare the data to several simple models. First, we increase the CO abundance by 20\% over the nominal value in the northern hemisphere and decrease it by 20\%  in the south, and vice versa. A southern hemisphere increase in the CO abundance is the better match of the two, but does not reproduce the brightening we observe near the south pole and causes an increase in $\chi^2$ for the 66 MHz, 213 MHz, and combined datasets. We also try three slightly more complex models. In the first case, we increase the CO abundance by 20\% at high latitudes (above 60$^\circ$), and decrease it by 20\% elsewhere. We repeat this for an increase at mid-latitudes (between 30 and 60$^\circ$N and S) and near the equator (30$^\circ$S to 30$^\circ$N). We find that an increase in the CO at mid-latitudes provides a qualitatively good match to the data  at 66 and 213 MHz and decreases the value of $\chi^2$ for the full dataset. However, for frequency offsets less than 66 MHz, the quality of the fit generally decreases. Despite this, we do not rule out a model in which the CO abundance is higher at mid-latitudes.

\subsubsection{Horizontally uniform CO, meridionally varying temperature}\label{sec:covartp}

Variations in the zonal mean temperature have been observed near Neptune's tropopause in infrared images from the {\it Voyager 2} spacecraft \citep{conrath98} and the Very Large Telescope (VLT) \citep{orton07}.   The latitude coverage of the {\it Voyager} IRIS data extend from 80$^\circ$S to 30$^\circ$N: \cite{conrath98} find a 100-mbar temperature minimum (4-5 K cooler than the equator) near 45$^\circ$S, as well as a decrease in temperature at the northernmost extent of their data.  The measurements of \cite{orton07} extend all the way to the south pole. These authors find similar latitudinal behavior of the temperature from their 17.6 (Orton model `A') and 18.7 $\mu$m VLT (Orton model `B') images. Near the south pole, they observe 100-mbar temperatures that are 7-10 K higher than elsewhere on the planet. 

To test the impact of these temperature variations on our models, we adjust the \cite{fletcher10} temperature profile with latitude to match the temperature variations reported by \cite{conrath98} and \cite{orton07} at 100 mbar. For simplicity, the temperature profile is modified by the same temperature offset at all altitudes above 1 bar. We acknowledge that this paradigm is most likely overly simplistic: the 2D temperature cross sections of  \cite{conrath98}  show that temperature variations appear to be located primarily near the tropopause,  with a gradual decrease in the meridional variations with altitude to nearly uniform within a few scale heights.  However, Fig. \ref{fig:mapcontrib} illustrates that much of the emission at  5-200 MHz from line center originates from near the tropopause, and from continuum opacity sources well below the tropopause (where the temperature should not vary and is not adjusted). We expect that models of this form are adequate for a first test of the effects of temperature variations on our maps. We find that these varying temperature models provide a good match to the data overall, and decrease the value of $\chi^2$ at all frequencies (Fig. \ref{fig:comparison} and Table \ref{tab:chisq}). The Orton models of the zonal mean temperature profile are in general a better match to the data than the \cite{conrath98} solution; this is partially because the  \cite{conrath98} model does not extend to the south pole. 

\begin{table*}[t!]
\scriptsize
\begin{center}
\setlength{\tabcolsep}{0.05in}

\caption{}{Calculated $\chi^2$ values for the latitude-binned data and a set of test models\label{tab:chisq}. See Section \ref{sec:covar} for a description of each model. Every other bin, starting with the bin centered at 80$^\circ$S, is used in calculated $\chi^2$, since data from neighboring bins are correlated. For individual maps (unscaled models), $DOF=4$ and the probability $p$ of obtaining $\chi^2>\chi^2_4$ due to chance is less than 0.05 for  $\chi^2_4>9.5$. For the scaled models at $\delta \nu=$ 213 MHz, $DOF=3$ and $p\leq 0.05$ for $\chi^2_3> 7.8$. When considering all frequency offsets simultaneously, $DOF=28$ and $p\leq 0.05$ for $\chi^2_{28}> 41.3$ ($DOF=27$ and $p\leq 0.05$ for $\chi^2_{27}> 40.1$ for all data when the $\delta \nu=$ 213 MHz data are scaled.) }
 \vspace{\baselineskip}
 
\begin{tabularx}{1.0\textwidth}{p{2.cm} p{1cm}  l p{.95cm} p{.95cm} p{.95cm} p{.95cm} p{.95cm} p{1.cm} p{.95cm}  p{.95cm} p{1.1cm} p{1.1cm} p{1.1cm} p{1.1cm} p{1.1cm}}
\hline
 Frequency offset & $\chi^2$ &  & & & & &  & & &  & & & &  \\
									 & nominal model & 50\% CO & 80\% CO &120\% CO & 150\% CO & +CO in N & +CO in S & +CO high lat & +CO mid-lat & +CO equator & Orton 2007 `A' & Orton 2007 `B' & Conrath 1998 \\
0.0 MHz  	&  1.9	& 2.9		& 0.9		& 5.9		& 18.8	& 1.6		& 4.3		& 2.0		& 2.9		& 1.3		& 1.2		& 1.2		& 0.9	\\
4.6 MHz  	& 2.8		&10.4	& 3.9		& 2.8 	& 4.5		& 4.6		& 2.2		& 3.7		& 3.1		& 3.6		& 1.2		& 1.2		& 1.3	\\
9.2 MHz  	& 2.0		&15.8	& 3.9		& 1.5		& 1.4		& 3.9		& 1.3		& 3.9		& 2.1		& 3.0 	& 0.9		& 0.8		& 1.5	\\
13.8 MHz 	& 2.8		&30.0	& 7.6		& 1.4 	& 1.4		& 6.5		& 1.7		& 7.6		& 4.2		& 3.7		& 1.9		& 1.8		& 2.5	\\
18.4 MHz 	& 2.1		&33.5	& 7.3		& 1.1		& 2.5		& 6.3		& 0.9		& 7.0		& 2.7		& 4.5		& 0.9		& 0.8		& 0.9	\\
66 MHz	&10.8	&206.5	& 33.0	& 29.6	& 97.5	& 31.7	& 14.6	& 34.0	& 3.0		& 37.5	& 4.3		& 2.6		& 3.1	\\
213 MHz	&10.2	&145.1	& 15.3	& 49.7	& 165.1	& 20.6	& 25.3	& 19.4	& 1.5		& 34.9	& 8.9		& 7.3		& 8.7	\\
\\
213 MHz, scaled&6.3&4.5\footnote{Best-fit scale factor is greater than 3\%, which is considered unlikely given the results for the continuum data.}	&5.7	&6.7$^\text{a}$	&6.9$^\text{a}$  	&19.4	&5.3	&14.2	&1.2	&31.2	&0.7	& 0.7	&2.2	\\
\\
all 		&32.6	& 444.2	&71.9	&91.9	&291.3	& 75.2	& 50.3	&77.5	& 19.5	&88.5	& 19.3	&15.7 & 18.9 \\
all,  213 MHz scaled &28.7& 303.6 	& 62.3	& 49.0	& 133.1	& 74.0	& 30.3 	& 72.2	& 19.2 	& 84.9	&  11.2	& 9.1 & 12.4 \\
\hline
\end{tabularx}
\end{center}

\end{table*}

The frequency at which the least improvement is observed is 213 MHz from line center. Figure \ref{fig:comparison} shows that the varying temperature models do qualitatively reproduce the latitudinal behavior seen at this frequency; however, there appears to be an overall offset between the data and these models. Of the wavelengths considered, the data at $\delta \nu =$213 MHz has the greatest contribution from continuum emission (Fig. \ref{fig:mapcontrib}), and therefore we consider the possibility that such an offset exists due to the continuum issues discussed in Section \ref{sec:continuum}. To test this, we refit each of the models for $\delta \nu =$ 213 MHz, allowing for an overall scaling of the models to best match the data.  We find that the fit improves dramatically for the varying temperature models when we allow for such a scale factor. We note that the fit also improves for several other models when we allow a scale factor in the fitting process; however we caution that in several instances the scale factors found in the fit have a value greater than 3\%, which seems unlikely given the results of Section \ref{sec:continuum}. These cases are indicated in Table \ref{tab:chisq}. 

To summarize, we find that the following models are not rejected by our $\chi^2$ analysis: the nominal model, in which the CO and thermal profiles are horizontally uniform;  the model in which the stratospheric CO abundance is increased at mid-latitudes and decreased elsewhere; and the three models in which the temperature profile varies with latitude to match the results of \cite{orton07} and \cite{conrath98}. Of these models, the nominal model appears to be ruled out as a fit to either the $\delta \nu =$66 MHz or the $\delta \nu =$213 MHz data when considered alone; furthermore, the four indicated test models all offer an improvement in the fit quality to the full dataset over the nominal model, with the greatest improvement for the varying temperature models. This improvement is even more dramatic when we allow the models at 213 MHz to have an applied scale factor to account for continuum opacity issues.

\section{Summary and conclusions}

Maps of Neptune are presented at frequency offsets of $\delta \nu=0$ MHz -- 5.6 GHz from the center of the CO (2-1) line at 230.538 GHz. Far from the CO line, the observed emission is primarily from the continuum, and the opacity is dominated by H$_2$S absorption and H$_2$ CIA. Our observations show that the region near the south pole is brighter than other latitudes. Similarly, centimeter-wavelength observations have shown that the south pole \citep{martin06,martin08,hofstadter08} and equator \citep{butler12} are bright compared to the rest of the planet. This has been attributed to a global circulation pattern in which air rises at mid- southern and northern latitudes and subsides near the equator and south pole; the subsidence of dry air would decrease the gas opacity allowing thermal emission from deeper levels to be observed. We model the potential effects of three opacity sources on Neptune's millimeter continuum:  the H$_2$S abundance,  the tropospheric methane abundance, and deviations from equilibrium in the {\it ortho/para} ratio of hydrogen. We find that our observations can be reproduced by latitudinal variations in the H$_2$S opacity; a good match  to the data is found by a model in which the H$_2$S opacity has the nominal value from 10$^\circ$N to 40$^\circ$N, 0.3 times the nominal value from 40$^\circ$S to 10$^\circ$N, and 0.03 times the nominal value between 90$^\circ$S and 40$^\circ$S. Such a decrease in the H$_2$S opacity near the south pole could cause an increase in the brightness temperature in this location at longer wavelengths as well.

For opacity sources that dominate at higher altitudes in the atmosphere, such as H$_2$ CIA, a smaller change in the opacity would be enough to produce the latitudinal trend in brightness observed in our continuum data. We can alternatively model the observed latitudinal variations by decreasing the CH$_4$ mole fraction from 0.044 at 10$^\circ$N -- 40$^\circ$N to 0.022 at 40$^\circ$S -- 10$^\circ$N and 0.0055 between 90$^\circ$S and 40$^\circ$S. While most previous studies \citep[e.g.][]{roe01,gibbard02,irwin11} assume that Neptune's methane mole fraction is constant in the troposphere with latitude, \cite{karkoschka11} find that CH$_4$ is depressed between 1.2 and 3.3 bar at high southern latitudes, compared to its abundance at  low latitudes. While we estimate somewhat larger variations in the CH$_4$ mole fraction (a total change of a factor of 8 rather than a factor of $\sim$3 found by \cite{karkoschka11}), the similarities between our estimated latitudinal CH$_4$ profile and that derived by \cite{karkoschka11} are conspicuous. Variations in the {\it ortho/para} ratio of hydrogen can also cause variations in the continuum intensity, and we find that a decrease from an equilibrium hydrogen fraction of 1.0 at the south pole to 0.6 at 40$^\circ$N produces the observed latitudinal trend. \cite{baines95} determined that the hydrogen in Neptune's upper troposphere is near equilibrium; however, \cite{conrath98} measured a decrease in the {\it para} hydrogen fraction from equilibrium at latitudes of  0$^\circ$ -- 60$^\circ$S, and an increase in the {\it para} fraction over the equilibrium value near the south pole and in the northern hemisphere. This result qualitatively agrees with the pattern we observe near the south pole and equator, although we do not see a rise in the intensity in the northern hemisphere in our data, which we would expect from an increase in the {\it para} fraction at these latitudes. We note that we do not investigate the effects of changing the adiabatic profile. This has been studied previously: \cite{depater93} show the effect of equilibrium, normal and intermediate hydrogen on Neptune's microwave spectrum between 0.1 and 10 mm by changing both the adiabatic profile and the CIA opacity.  They show that the brightness temperature is highest for intermediate H$_2$, and lowest for equilibrium H$_2$.

At frequency offsets $\delta \nu=$0--213 MHz from line center, where we are sensitive to the CO abundance and temperature in the lower stratosphere, we find that the south pole and equator are relatively bright compared to the region near 45$^\circ$S and 40$^\circ$N. Maps at $\delta \nu =$66 MHz or the $\delta \nu =$213 MHz are inconsistent with the nominal model, which has a horizontally uniform CO and thermal profile. Introducing the temperature variations derived by \cite{conrath98} or \cite{orton07} into our models successfully reproduces the observed latitudinal variations. As discussed by \cite{conrath98} and \cite{martin08}, the low temperatures at mid-latitudes are consistent with upward motion and adiabatic cooling, as suggested by the centimeter and millimeter continuum data. The increase in temperature near the south pole is likely due to increased solar insolence in this region \citep{hammel07b,orton07} and/or subsidence and adiabatic heating \citep{martin08}. We also compare the data to models in which the CO abundance varies with latitude; we find that a model in which the stratospheric CO abundance is increased at mid-latitudes and decreased elsewhere is also consistent with the data. We expect that any variations in the CO abundance must be less than 20\%, at the resolution of our observations. 

These data suggest the utility of measurements in and near CO rotation lines for constraining latitudinal variations in the thermal profile and opacity in Neptune's lower stratosphere and upper troposphere. Future observations can improve on these results in several ways. First of all, poor weather during our observations increased the noise level and introduced ripples into the final maps, limiting the power of the dataset for ruling out alternative models. Secondly, in addition to being interesting in its own right, a better characterization of Neptune's continuum opacity is critical to better modeling data within the CO line. For example, our analysis from \cite{luszcz13} indicates that H$_2$S opacity has a small but non-negligible effect on the retrieved vertical CO profile. Finally, in order to disentangle the effect of temperature and CO variations, spatial maps of several CO transitions are required.  The Atacama Large Millimeter/sub-millimeter Array (ALMA), which is located  on the Chajnantor plateau at an altitude of 5000m and will consist of 66 7- and 12-m antennas when completed, will provide unprecedented sensitivity and resolution for such observations.

\section*{acknowledgements}
The data presented in this work were obtained with CARMA. Support for CARMA construction was derived from the states of California, Illinois, and Maryland, the James S. McDonnell Foundation, the Gordon and Betty Moore Foundation, the Kenneth T. and Eileen L. Norris Foundation, the University of Chicago, the Associates of the California Institute of Technology, and the National Science Foundation. Ongoing CARMA development and operations are supported by the National Science Foundation under a cooperative agreement, and by the CARMA partner universities. This work was supported by NASA Headquarters under the NASA Earth and Space Science Fellowship program - Grant NNX10AT17H; and by NSF Grant AST-0908575. The authors would like to thank L. Fletcher for providing his temperature and CH$_4$ profiles, and G. Orton for providing his H$_2$ CIA absorption coefficients. The authors would also like to acknowledge R.L. Plambeck and A. Bauermeister for many helpful discussions. 


\newpage
\footnotesize

\bibliographystyle{apalike}
\bibliography{thesis}

\appendix
\section{Data reduction}\label{sec:reduction}
We perform editing and calibration on the raw visibility data using the Multichannel Image Reconstruction, Image Analysis and Display \citep[MIRIAD;][] {sault11} software package. We flag 10 edge channels in the narrow band centered on the CO line and three edge channels in the wide bands; the narrow window in the opposite sideband from the CO line is flagged entirely. Poor quality data are also flagged. After performing passband calibration, we correct for atmospheric and instrumental effects on the observed visibilities by determining the time-dependent antenna-based gains using the wide-band data. An initial self calibration is performed in two steps using the phase calibrator: a record-by-record phase solution is found for the phase calibrator to remove short-term variations. This is followed by a phase and amplitude self calibration on the phase calibrator using a time interval corresponding to one observing cycle; these gain solutions are then applied to the Neptune data to correct for slow variations in the antenna gains. Finally, a record-by-record phase-only self calibration is performed on Neptune itself to remove short-term phase variations in the Neptune data. We find that since the rms path errors are a large fraction of the wavelength of the observations, the phase coherence is too poor for self calibration to successfully determine the phase solutions for isolated remote antennas. This ultimately causes systematic errors in the final maps; we discuss this further in Appendix \ref{app:error}. 

\begin{figure}[htb!]
\begin{center}$
\includegraphics[width=0.9\textwidth]{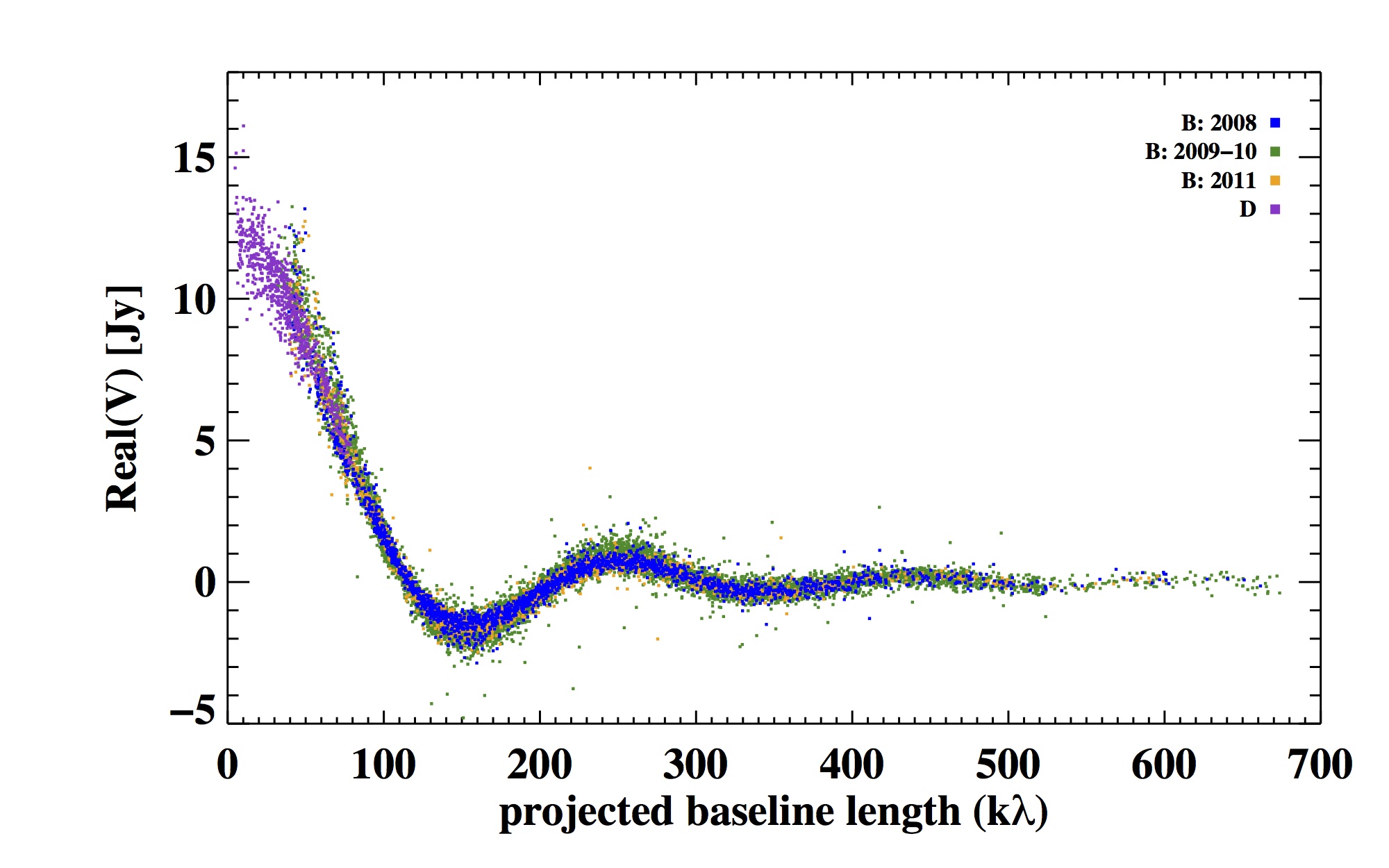}
$
\end{center}
\caption[Real part of visibility amplitude vs. (u,v) distance]{\label{fig:uv} The real part of the visibilities, shown as a function of projected baseline length. Each data point is the average over a single observing cycle (8 minutes in the B configuration, 15 minutes in the D configuration). The purple points are the short baseline, D-array data, which sample large angular scales. The B-array data (blue/green/orange for data taken in the winter of 2008/2009--2010/2011, respectively)  measure smaller-scale structure. Both arrays have some baselines with projected lengths of 40-80 k$\lambda$; the D-array data at these baseline lengths are used to flux-calibrate the B-array data as described in Appendix \ref{sec:reduction}. }
\end{figure}

The absolute flux scale of CARMA visibility data is typically set using observations of a calibrator of `known' flux, usually a planet. Flux calibration using this method is expected to be accurate to roughly 20\%. During our D-array observations, only a single dataset contained observations of a primary flux calibrator (MWC349\footnote{\url{http://cedarflat.mmarray.org/fluxcal/primary\_sp\_index.htm}}). As described in \cite{luszcz13} we flux-calibrated this dataset in the standard way using an assumed value of 1.86 Jy for MWC349 at 230 GHz. Then, we binned the Neptune visibility data for each D-array dataset into 8 (u,v) bins between 0 and 80 k$\lambda$, and adjusted the flux calibration to align the binned visibilities between tracks as much as possible. Finally, during the analysis of Neptune's disk-integrated spectrum, we determined an overall correction factor to the flux calibration that optimizes the match between the data and our best-fit model. As discussed in \cite{luszcz13}, the continuum slope of our best-fit model is not a perfect match to the data, resulting in model fluxes that are systematically too high by $\sim 2\%$ near 225 GHz and too low by $\sim 2\%$ near 236 GHz as compared to the disk-integrated measurements. For the present analysis, we scale the D-array visibilities by the flux correction factor determined from the fit to our nominal model (Section \ref{sec:comodel}).  We then bin the Neptune visibility data for each individual B-array dataset in intervals of 10 k$\lambda$ at (u,v) distances for which both  B- and D-array data are present (40--80 k$\lambda$). Finally, we adjust the overall flux scale for each B-array track to align the binned visibility data with the D-array data. The real parts of the calibrated visibilities from the B and D configurations are plotted as a function of projected baseline length in Fig. \ref{fig:uv}, where each point is the average over one observing cycle. For reference, the shape of the visibility function corresponding to a circularly symmetric, uniform disk is a Bessel function. After flux calibration, Neptune's measured disk-integrated flux density is 12.3 Jy at 225 GHz and 12.2 Jy at 229.5 GHz. 

\section{Imaging and deconvolution}\label{app:imagedecon}
We produce maps of Neptune's flux in four steps: first we convert our visibility data into a set of maps of the sky intensity distribution. Then, we estimate the noise in the maps due to atmospheric and instrumental effects. Using this estimate, we perform deconvolution to remove the response to source structure in the side lobes of the synthesized beam. Finally, we average images to increase signal-to-noise, and bin the data according to latitude to look for trends in Neptune's observed brightness. 

\subsection{Imaging}\label{app:image}
We use the Common Astronomy Software Applications (CASA) data reduction package\footnote{http://casa.nrao.edu} to convert our visibility data into maps of the sky intensity distribution. The frequency resolution of these maps is selected to be lower than the original channel spacing of the visibility data in order to improve the signal-to-noise in the images: from the 62 MHz band centered on the CO line, we define 9 frequency intervals of width 4.6 MHz. From the 500 MHz bands, we select eight 125 MHz frequency windows for mapping. The original channels are combined in the imaging step using multi-frequency synthesis mode. As mentioned in Section \ref{sec:obs}, the imaging windows are chosen to cover the frequencies where we have the most data. The frequencies of these imaging windows are listed in Table \ref{tab:image1}. 

We apply intermediate weighting to the visibilities with a Briggs visibility weighting robustness parameter \citep{briggs95} of 0.0. We use the mosaic option to properly handle the three different primary beam patterns of the heterogeneous CARMA array. The pixel size used in the maps is 0.09''; the shape of the synthesized beam varies between images, with a FWHM in the range of 0.33--0.37'' x 0.37--0.39''. 

\subsection{Error determination}\label{app:error}
Calibration errors and atmospheric fluctuations introduce complex-valued multiplicative errors into the data. As noted in Section \ref{sec:obs}, the rms path errors in our data are as high as 350 $\mu$m, which is roughly $1/3$ of the observing wavelength and well above CARMA's 150 $\mu$m criterion for good 1-mm observing conditions. We are able to correct for these errors on some baselines using self calibration (Appendix \ref{sec:reduction}); however the phase coherence is too poor for this technique to be successful on isolated remote antennas, which results in increased errors in our final maps above the value expected from thermal noise from the atmosphere and receiver system. 

We can estimate the noise by subtracting a model from the data and making an image. The average rms in the 4.6 MHz model-subtracted images is 28 mJy/beam, 2.7 times the value expected from thermal noise. The rms in the 125 MHz images is on average $7.8$ mJy/beam, 3.3 times the value expected from thermal noise. 

Another estimate of the noise can be be obtained by making an imaginary image from the non-Hermitian part of the data. For a real valued sky brightness intensity, the visibility function is Hermitian. An image of the non-Hermitian part of the data is a direct measure of the errors in the visibility data. The imaginary images show Gaussian distributions of pixel values with rms values that are in agreement with the noise estimate found by the previous method. The noise estimates from these two methods are presented for the individual maps in Table \ref{tab:image1}.

\begin{table}[htb!]

\scriptsize
\begin{center}
\setlength{\tabcolsep}{0.05in}

\caption{}{Imaging frequencies, listed in order of increasing distance from line center. The imaging frequency resolution is coarser than the original channel spacing of the data. The image noise is estimated in two ways: method 1 involves measuring the rms in an image made from model-subtracted data. Method 2 is to find the rms in an imaginary image made from the data. \label{tab:image1}}

 \vspace{\baselineskip}
 \footnotesize
\begin{tabularx}{.7\textwidth}{l p{2.1cm} p{2.3cm} p{2.3cm} p{2.3cm} }
\hline
 Center frequency & Frequency offset from line center & Image frequency width &RMS, method 1 (mJy/beam)	&RMS, method 2 (mJy/beam) \\
230.538 GHz &   0.00 MHz & 4.6 MHz		&28.2	&28.3\\
230.533 GHz &  $-4.62$ MHz & 4.6 MHz	&28.6	&28.5\\
230.543 GHz &   4.62 MHz & 4.6 MHz		&27.8	&27.8\\
230.529 GHz &  -9.23 MHz & 4.6 MHz		&27.9	&28.2\\
230.547 GHz &   9.23 MHz & 4.6 MHz		&28.7	&28.8\\
230.524 GHz & $-13.8$ MHz & 4.6 MHz	&26.7	&26.6\\ 
230.552 GHz &  13.9 MHz & 4.6 MHz		&29.1	&29.0\\
230.520 GHz & $-18.4$ MHz& 4.6 MHz		&27.7	&27.5\\
230.556 GHz &  18.5 MHz & 4.6 MHz		&26.9	&26.7\\
230.604 GHz &  66.0 MHz& 125 MHz		&8.13	&8.01\\
230.751 GHz &   213 MHz& 125 MHz		&7.64	&7.59\\
229.601 GHz &$-0.937$ GHz& 125 MHz	&8.39	&8.43\\
229.476 GHz & $-1.06$ GHz& 125 MHz		&8.57	&8.53\\
225.600 GHz & $-4.94$ GHz& 125 MHz		&7.58	&7.59\\
225.475 GHz & $-5.06$ GHz& 125 MHz		&7.21	&7.21\\
224.450 GHz & $-6.09$ GHz& 125 MHz		&7.41	&7.44\\
224.325 GHz & $-6.21$ GHz& 125 MHz		&7.44	&7.37\\
\hline
\end{tabularx}
\end{center}
\end{table}

\subsection{Deconvolution}
The imaging step is followed by deconvolution, which attempts to reconstruct the true sky brightness distribution from the observed image using the synthesized beam. Two different deconvolution algorithms are commonly used: an iterative point source subtraction algorithm, \CLEAN\, which is well-matched for deconvolving compact source structures, and maximum entropy, a gradient search algorithm, which maximizes the fit to an a-priori image, in a least squares fit to the (u,v) data. After investigating both deconvolution strategies  and several variations of \CLEAN\ (Appendix \ref{app:decon}), we deconvolve our synthesized maps using the Clark \CLEAN\ algorithm \citep{clark80} with a gain factor of 0.05, cleaning down to the rms noise level of  the image. A starting \CLEAN\ model is provided based on the radiative transfer solution for the nominal CO profile from \cite{luszcz13} (Section \ref{sec:comodel}). We note that the resulting maps do not appear to be affected by the details of the starting \CLEAN\ model. For a more complete comparison of the imaging and deconvolution strategies tried, see Appendix \ref{app:decon}.

\subsection{Averaging}

After the initial mapping procedure, we average deconvolved images where possible to further improve the signal-to-noise: the contribution function (Fig. \ref{fig:mapcontrib}) shows that at an offset of 5 GHz from line center, the maps are almost entirely due to continuum emission from pressures of 1-5 bar. We therefore average the four maps at offsets of 5-6 GHz from line center to produce a single continuum image (Fig. \ref{fig:avgcont}). We also combine the maps at 229.48 and 229.60 GHz (1.04 and 0.96 GHz from line center, respectively), giving us  three high signal-to-noise images at offsets of 66 MHz, 213 MHz and 1 GHz from line center (Fig. \ref{fig:avgwide}). In the core of the CO line where we have an imaging resolution of 4.6 MHz, we average maps with the same absolute frequency offset from line center (Fig. \ref{fig:avgnar}). The contribution functions for these image pairs are very similar, so this allows us to increase our signal to noise while maintaining sensitivity to emission from different atmospheric pressures.

The noise in each of these averaged maps is approximated as the root mean square sum of the errors in the original maps, divided by the square root of the number of input images. The higher of the two noise estimates for each input image (Table \ref{tab:image1}) is used. We note that averaging of images takes place after cleaning, which is a non-linear process. We also caution that systematic errors, for example those caused by gain calibration errors, will not decrease with more averaging. This can cause our errors to underestimate the true noise in the images. Therefore, the noise values derived for the final, averaged maps should be taken as best estimates. These values are presented in Table \ref{tab:mapfreq}.

Longitudinal brightness variations due to changes in the atmospheric properties should be averaged out in the long data integrations, but latitudinal variations may reflect meridional trends in the composition and/or temperature. To look for meaningful trends and further increase signal to noise, we bin the data in each map in bins of latitude. We use the ephemeris data from JPL Horizons\footnote{\url{http://ssd.jpl.nasa.gov/?horizons}} to get the value of the latitude that each pixel location in the image would have if it were not convolved with the synthesized beam. To determine the variation of the intensity in the maps with physical location on the planet, we select 7 non-overlapping latitude bins; the width of each bin is approximately equal to one beam, such that data separated by at least one bin will be largely uncorrelated. For each latitude bin, we make a model mask: a pixel in the mask is set to 1 if the latitude at that pixel location is within the latitude bin range; otherwise it is set to 0.  We then convolve the mask with the synthesized beam, and multiply this mask by the data. The total of this masked image divided by the number of pixels in the mask gives a weighted average of the data over the desired latitude range.  The error for each binned point is estimated as the image noise in mJy/beam divided by the square root of the number of beams within the binned region, which varies from just over 1 beam for the bins centered at 80$^\circ$S and 40$^\circ$N, to nearly 6 beams in the bin centered at 20$^\circ$S. For direct comparison with the models, we convolve the models with the \CLEAN\ beam and multiply by the same masks.

\section{Comparison of deconvolution techniques}\label{app:decon}
\begin{figure}[htb!]
\begin{center}$
\includegraphics[width=0.5\textwidth]{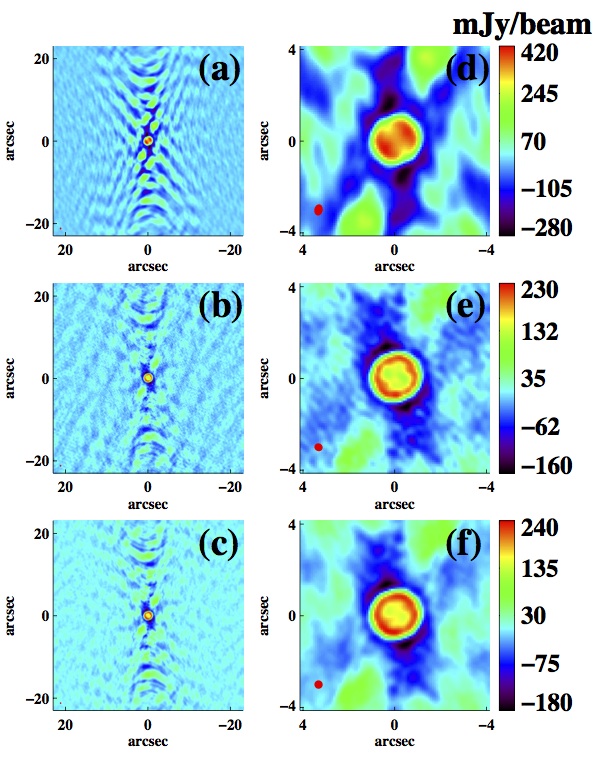}
$
\end{center}
\caption[Synthesized images, B array only]{\label{fig:S4dirty}   Synthesized images, B-array (long baseline) data only. Made using natural weighting (top), uniform weighting (middle) and robust weighting with a Briggs visibility weighting robustness parameter \citep{briggs95} of 0.0 (bottom). On the left we show the full maps; on the right we zoom in on the planet. The beam is indicated by the red oval in the bottom left corner of each image. }
\end{figure}

\begin{figure}[htb!]
\begin{center}$
\includegraphics[width=0.5\textwidth]{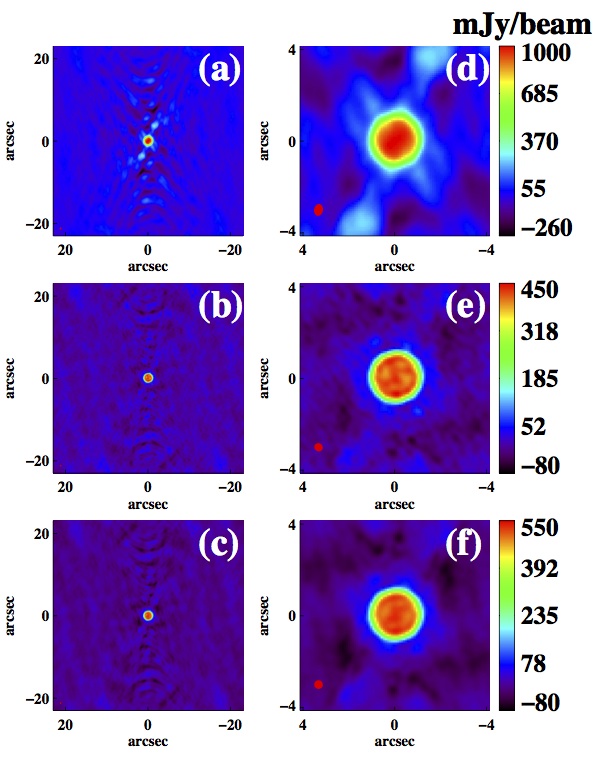}
$
\end{center}
\caption[Synthesized images, B+D arrays]{\label{fig:S3dirty}   Synthesized images, using all data (B and D arrays). Made using natural weighting (top), uniform weighting (middle) and robust weighting with a Briggs visibility weighting robustness parameter \citep{briggs95} of 0.0 (bottom). On the left we show the full maps; on the right we zoom in on the planet. The beam is indicated by the red oval in the bottom left corner of each image. }
\end{figure}

Neptune, with its smooth, bright disk and sharply defined edges, presents a challenge for imaging and deconvolution. In order to evaluate the effect and significance of our weighting function and deconvolution technique, we perform a series of comparisons using a subset of our data that spans 437.5 MHz in frequency, centered at 225.74 GHz (nearly 5 GHz from the center of the CO (2-1) line). These data represent an instance where there is overlap between the D-array data and all epochs of B-array data. To determine the importance of the short-spacing (D-array) data in the final maps, we imaged the B-array data separately, in addition to imaging the full combined data set. The synthesized maps from both data subsets are shown in Figs. \ref{fig:S4dirty} and \ref{fig:S3dirty}, with three different weighting functions (natural, uniform and intermediate) applied. Natural weighting (top), which weights each visibility by the inverse of its noise variance, gives the best sensitivity, at the expense of the shape of the synthesized beam and sidelobe levels. Uniform weighting (middle) adjusts the weight of each visibility so that the density of visibilities is uniform across the (u,v) plane. This minimizes sidelobe levels, but increases the noise level in the map. Robust weighting with a Briggs visibility weighting robustness parameter \citep{briggs95} of 0.0 (bottom) is a good compromise between the two; maintaining a well-behaved beam shape with less increase to the noise level.

\begin{figure}[htb!]
\begin{center}$
\includegraphics[width=0.9\textwidth]{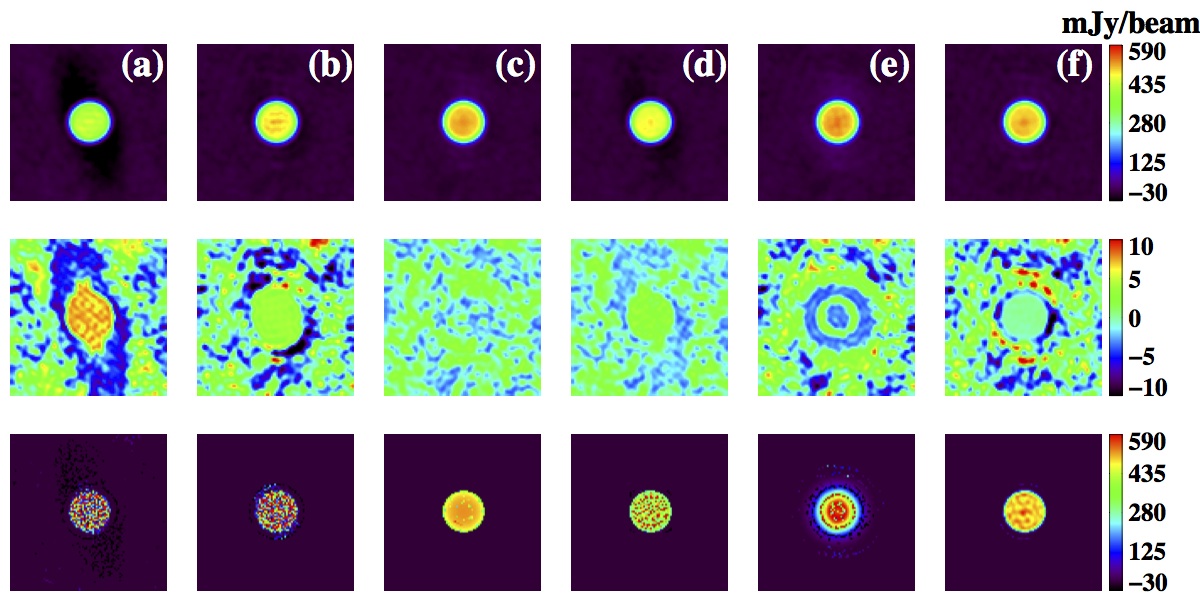}
$
\end{center}
\caption[Comparison of deconvolution techniques, B-array data]{\label{fig:S4cmap}   Comparison of deconvolution techniques, for the B-array data only. Discussion of techniques (a)-(f) is given in Appendix \ref{app:decon}. For each technique, the deconvolved map (top) residual map (middle) and deconvolution model (bottom) are shown. The image size is the same as in Fig. \ref{fig:S3dirty}, right. The deconvolution model is the model of the sky that includes any input  (starting) model component plus the components determined by the deconvolution. The deconvolved map is the synthesized map, less the model convolved by the synthesized beam, plus the model convolved by a gaussian fit to the synthesized beam. The residual map is the synthesized map, less the model convolved by the synthesized beam.}

\end{figure}

\begin{figure}[htb!]
\begin{center}$
\includegraphics[width=0.4\textwidth]{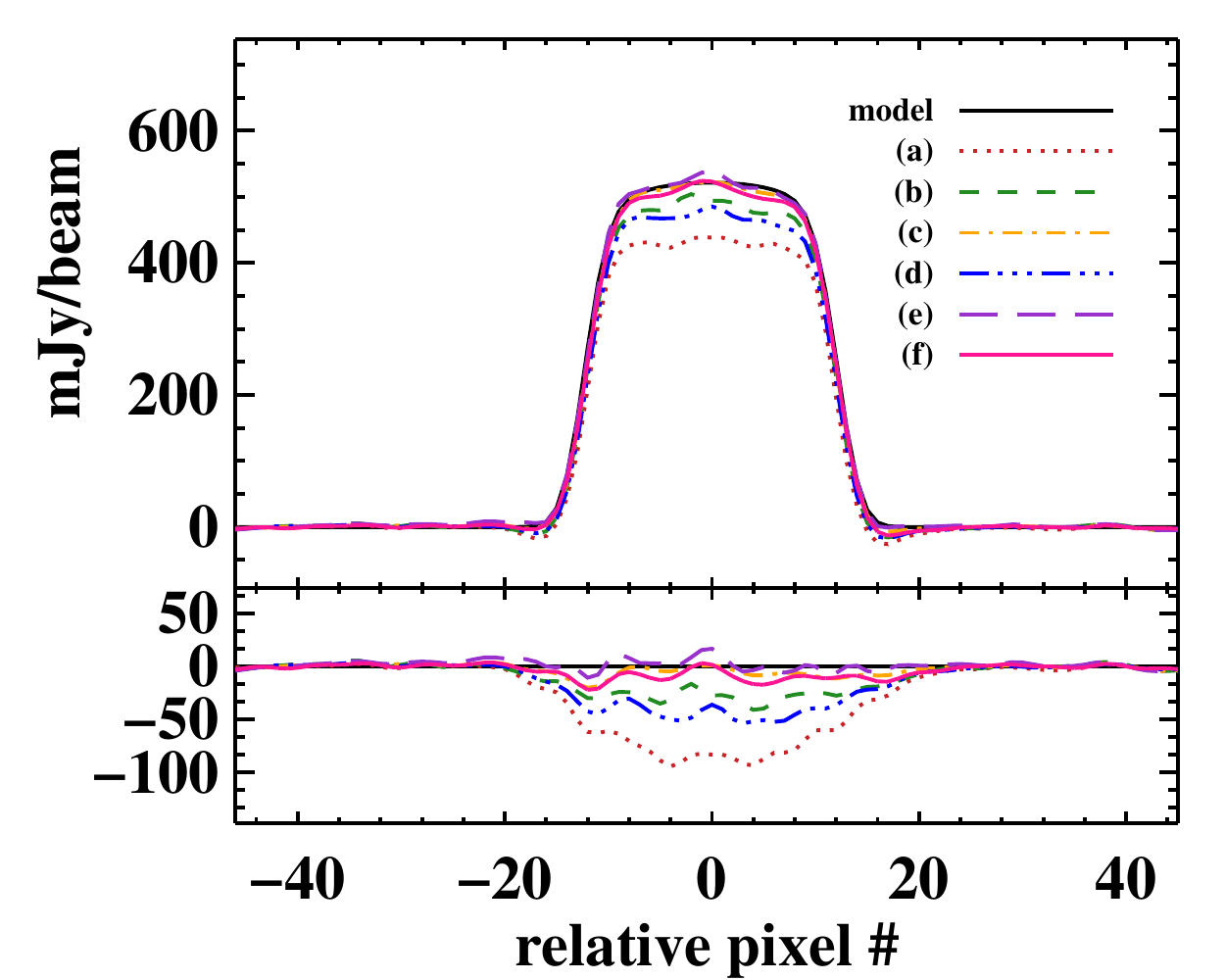}
$
\end{center}
\caption[Slice through clean maps, B-array data]{\label{fig:S4slice}   Comparison of maps from deconvolution techniques (a)-(f), for the B-array data only. Plots are made by slicing horizontally through the center of the maps, and shown as relative pixel numbers, where the center pixel is at 0 (each pixel has a width of 0.09''). Also shown is our nominal model (black) for the data. In the bottom plot, we subtract the nominal model from each of the map slices.  }
\vspace{\baselineskip}

\end{figure}

\begin{figure}[htb!]
\begin{center}$
\includegraphics[width=0.9\textwidth]{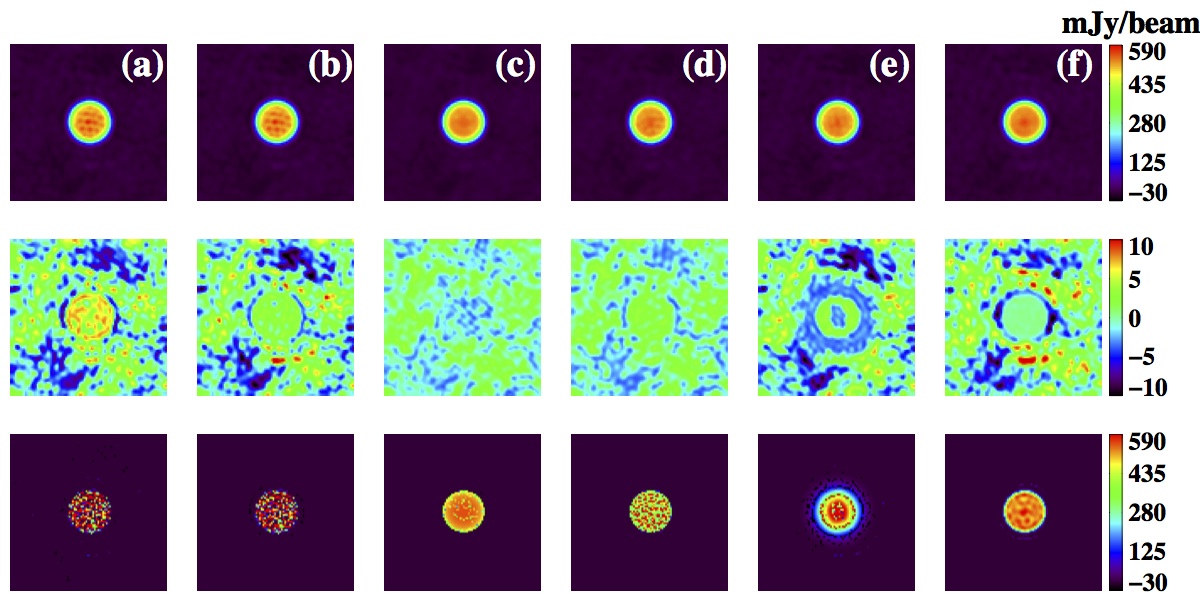}
$
\end{center}
\caption[Comparison of deconvolution techniques, B+D array data]{\label{fig:S3cmap}   Same as Fig. \ref{fig:S4cmap}, except for all data (B and D arrays). }
\end{figure}

\begin{figure}[htb!]
\begin{center}$
\includegraphics[width=0.4\textwidth]{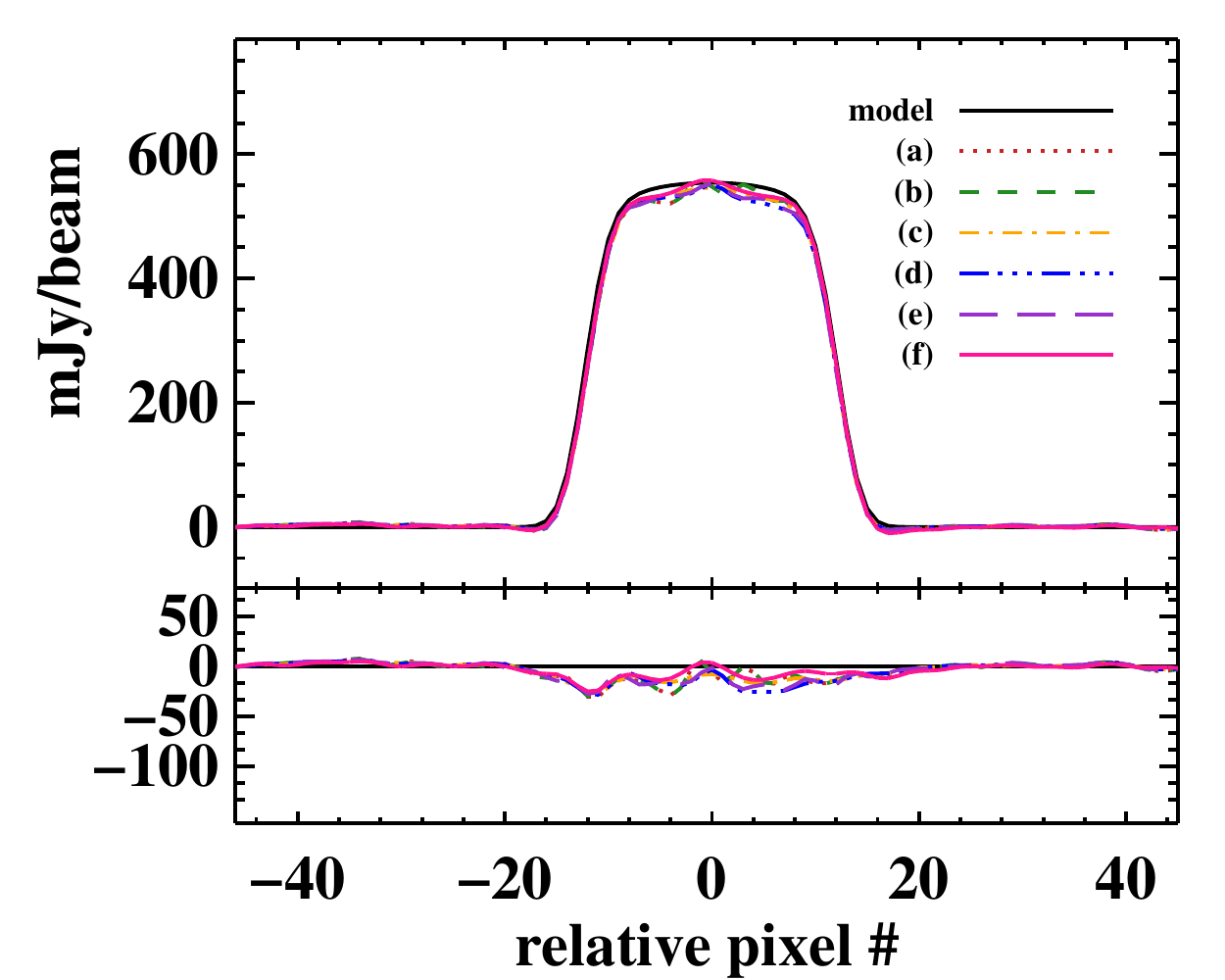}
$
\end{center}
\caption[Slice through clean maps, B+D array data]{\label{fig:S3slice}   Same as Fig. \ref{fig:S4slice}, except for all data (B and D arrays). }
\vspace{\baselineskip}

\end{figure}

We experimented with several deconvolution techniques; six of them are presented here. The results are summarized in Table \ref{tab:clean} and shown in Figs. \ref{fig:S4cmap} - \ref{fig:S3slice}.

\begin{table}[htb!]
 \scriptsize
\begin{center}
\setlength{\tabcolsep}{0.05in}

   \caption[Comparison of deconvolution strategies]{}{ Comparison of deconvolution strategies. \label{tab:clean}}
 \vspace{\baselineskip}
   
\begin{tabular}{l l l l l l l l l l l}

\hline
		&			&				&		&				& \multicolumn{3}{l}{B-array data only}	& \multicolumn{3}{l}{Full dataset}\\
	 	&Algorithm	& Region	&Input  model 	&Threshold	&Flux\footnote{Total flux recovered; nominal clean model gives a total flux of 12.740 Jy}  &RMS\footnote{standard deviation of the full residual map} & NITER\footnote{number of \CLEAN\ or maximum entropy iterations performed}& Flux$^a$  & RMS  & NITER$^c$\\
 	&	&	&  	& (mJy/beam)	& (Jy) &  (mJy/beam) &  &(Jy)  & (mJy/beam)  &   \\
(a)		& CLEAN		& -				&-		& 8.6		&	9.10 	&2.1	&6810	&12.26	& 2.0		&2180\\
(b)		& CLEAN		& 1.5''  circle		&-		& 4.3		&    11.56	&2.0	&8451	&12.25	& 2.0		&2579\\
(c)		& CLEAN		&	square\footnote{square extends just beyond the Neptune model disk- approximately 2.3'' on a side}													& nominal&4.3 		&    12.62  	&1.5	&6477	&12.24	& 1.4		&9918\\
(d)		& CLEAN		&       square$^d$& flat	& 4.3		&    10.88    	&1.4	&11344     &12.24	& 1.6		&8168	\\
(e)		& MSCLEAN	&2.5'' \ circle			&-		& 4.3     		&    13.34	&1.9	&680	&12.24	& 2.0		&628	\\
(f)		& Max. entropy	&1.5'' \ circle		&-		& 4.3 		&    12.44	&5.5	&88	&12.50	& 6.2		&71	\\
\hline
\end{tabular}
\end{center}
\vspace{\baselineskip}

\end{table}

\footnotesize

f

\vspace{\baselineskip}

\begin{enumerate}[(a)]
\item Classical \CLEAN\, no clean region: We used the CASA implementation of the Clark \CLEAN\ routine with a gain factor of 0.05. A maximum of 100,000 iterations were permitted, unless a clean threshold of 8.6 mJy/beam was reached. This threshold was selected to be twice the measured rms for the synthesized map. We find that a significant amount of noise is cleaned, particularly in the case of the B-array data subset, for which a strong negative ``clean bowl" is present due to the missing short spacing information. As a result, the residual map is artificially low, and only 3/4 of the total flux from the planet is recovered when only B-array data are included. 

\item Classical \CLEAN, inside a circle of radius 1.5'': We restrict \CLEAN\ to a circular region of radius 1.5'', just outside of the planet. Since most of the flux within this region is source (not noise), we use a lower clean threshold of 4.3 mJy, equal to the measured rms of the map.  This technique recovers  more than 95\% of the expected (model) flux from Neptune, even when the D-array data are not included. However, the noise outside of the \CLEAN\ region is higher than in many of the other test methods, and the disk appears very `lumpy' in the final \CLEAN\ map, presumably because the flux is being reproduced by a set of point sources, which is unrealistic in the case of Neptune.

\item Classical \CLEAN, starting with best-guess clean model: \CLEAN\ is given a starting model, which is taken from our radiative transfer model described in Section \ref{sec:comodel}. The input model uses 0.005'' pixels that cover a square region that extends just beyond Neptune's limb. \CLEAN\ then proceeds in the same way as before, adding additional point source components (both positive and negative) within the spatial range the starting model until the \CLEAN\ threshold is reached. In general, we find this approach gives a good result: the noise is low across the map (even beyond the \CLEAN\ region), and the disk appears far smoother than for the previous two techniques. 

\item Classical \CLEAN, starting with a flat clean model: To address the possibility that our choice of input model has an unexpected effect on the output map (perhaps forcing the data to match our nominal model), we repeat case (c) using a flat input model, which is a disk of constant intensity, equal to the lowest intensity value of the best guess model in (c). We find that  without including the D-array data, some of the negative bowl from the missing short-spacing data remains with this technique, so that the measured flux from Neptune is low. However, when the D-array data are included, a flat input starting model produces a very similar map to the more detailed model in (c), and the noise properties of the map are similar for the two starting model cases. 

\item Multiscale \CLEAN:  As an alternative to classical \CLEAN\, we try the CASA implementation of multi-scale \CLEAN. Rather than modeling the sky brightness with a set of point-sources, multi-scale \CLEAN\ models the sky with components of several different size scales. For a detailed comparison of the multi-scale \CLEAN\ and traditional \CLEAN\ algorithms, see \cite{rich08}. We use a gain of 0.3; the threshold is set to 4.3 mJy. Five size scales are specified, ranging from a point source up to the diameter of Neptune. Since multi-scale \CLEAN\ will only place model components such that they are entirely contained within the cleaning region, we specify a larger cleaning radius of 2.5''. We find that multi-scale \CLEAN\ is able to recover Neptune's flux even when the short spacing data were omitted. Multi-scale \CLEAN\ also requires significantly fewer iterations to converge than classical \CLEAN\ , and agrees well with (c) without depending at all on an input \CLEAN\ model. However, the noise level outside of the cleaning region is somewhat higher than in (c) and (d). 

\item Maximum Entropy: We use the {\tt mosmem} routine in MIRIAD, which is an implementation of the Maximum Entropy Method (MEM). This algorithm, which is an alternative to \CLEAN\, produces a smooth positive image which maximizes the fit to an a-priori image, in a least squares fit to the (u,v) data. We restrict the deconvolution to a circle of radius 1.5'', and specify the total flux based on the expected value from the model. We allow for as many as 5000 iterations to reduce the image residuals to 4.3 mJy/beam. We found that  {\tt mosmem} converged quickly; however, this approach had the highest residual level of any of the methods we tried.

\end{enumerate}
From this comparison, we conclude that when the D-array (short baseline) data are included, there is no significant difference in the final maps from the different deconvolution methods. In particular, while a starting input \CLEAN\ model appears to improve the appearance of the final maps, it does not artificially force the resulting maps to agree with the starting model, as long as the short baseline data are included. Table \ref{tab:clean} lists the total flux densities in the maps. The maps produced using the full dataset produce a final flux density that is consistently lower than the model predicts. This result is consistent with the disk-integrated spectrum described in \cite{luszcz13}, which indicates that at 225 GHz the model over-predicts  Neptune's total flux.  When we image only the B-array data, the use of a starting model is more influential, reducing the \CLEAN\ bowl and increasing the total flux. 

As discussed in Appendix \ref{app:error}, calibration errors and atmospheric fluctuations introduce amplitude and phase errors into the visibility data. The `ringing' in the final maps is not an artifact of deconvolution, but is a result of these errors. This is supported by the fact that the rippling is observed in maps produced using both \CLEAN\ and Maximum Entropy deconvolution methods, and is in fact observable in the synthesized images (prior to deconvolution). These ripples have an amplitude as high as 1 mJy/pixel, and the average deviation between map (c) and a smooth model (scaled to the data amplitude) is 5.7 mJy/beam. 

Since the maps are not overly influenced by the inclusion of an input model, the appearance of map (c) is best, and the residuals in case (c) are the lowest, we select method (c) for our image deconvolution.


\end{document}